\documentclass[aps,prd,twocolumn,nofootinbib,superscriptaddress,floatfix]{revtex4-2}

\usepackage{graphicx}
\usepackage{amsmath,amssymb}
\usepackage{bm}
\usepackage{mathrsfs}
\usepackage{booktabs}
\graphicspath{{../figures/}{figures/}}

\newcommand{\safeincludegraphics}[2][]{%
  \includegraphics[#1]{#2}%
}

\begin{document}

\title{Core-Halo Mass Relation in Cosmological Vector Dark Matter}

\author{Jiajun Chen}
\email{chenjiajun@swu.edu.cn}
\affiliation{School of Physical Science and Technology, Southwest University, Chongqing 400715, China}

\author{Yonghao Yao}
\email{jianghuyinshi@email.swu.edu.cn}
\affiliation{School of Physical Science and Technology, Southwest University, Chongqing 400715, China}

\author{David J. E. Marsh}
\email{david.j.marsh@kcl.ac.uk}
\affiliation{Theoretical Particle Physics and Cosmology, King's College London, Strand, London WC2R 2LS, United Kingdom}

\begin{abstract}
We study the cosmological core-halo relation in vector dark matter using
three-component Schr\"odinger-Poisson simulations.  Starting from
cosmological vector-field initial conditions, which due to the evolution of the vector field during inflation are enhanced on small scales, we find that nonlinear
evolution begins almost immediately following matter-radiation equality and produces compact self-gravitating Proca-star condensates at the
centers of halos.  After confirming the central condensates through their
radial density profiles, we find the empirical relation
\(\widetilde M_\star\propto
\widetilde M_{\rm h}^{0.6403}\) between the Proca star mass and halo mass, although interestingly we find that we are only able to confirm Proca stars in $\mathcal{O}(10\%)$ of halos. This serves as important input for future studies of the abundance and merger rates of Proca stars in models of vector dark matter. We also examine
the vector-field structure of the objects through global longitudinal and
transverse polarization fractions and local spin density inside
halos, which increases over cosmic time.
\end{abstract}

\maketitle

\section{Introduction}
\label{sec:introduction}

The composition of dark matter is one of the most important unresolved
problems in modern cosmology.  Cosmological observations indicate that dark
matter accounts for a substantial fraction of the total energy density of the
Universe \cite{Aghanim:2018eyx}.  Among the many proposed candidates,
ultralight bosonic fields provide a well-motivated class of dark-matter
models~\cite{Kimball:2023vxk,Marsh:2024ury}.

The most widely studied examples are scalar or pseudoscalar ultralight fields,
including the QCD axion and axion-like particles
\cite{Dine:1982ah,Preskill:1982cy,Abbott:1982af,Widrow:1993qq,Suarez:2013iw,2015PhRvD..92j3513G,Uhlemann:2014npa,Hui:2016ltb}.
In the following, we refer to such models as scalar dark matter.  Scalar dark
matter can form compact self-gravitating solitons, or boson stars, whose
stability is supported by the balance between gravity and gradient energy
\cite{Kaup:1968zz,Ruffini:1969qy,Colpi:1986ye,Seidel:1993zk,Liebling:2012fv,Schive:2014dra,PhysRevLett.113.261302}.
In cosmological simulations, these solitonic cores appear at the centers of
virialized wave-dark-matter halos.  Their formation has been discussed both as
a consequence of coherence inherited from the initial collapse and as a result
of gravitational Bose-Einstein condensation in the kinetic regime
\cite{Schive:2014dra,Eggemeier:2019jsu,Levkov:2018kau,Chavanis:2020upb,Jain:2023ojg}.

A key phenomenological result of scalar wave-dark-matter simulations is the
core-halo relation, which connects the mass of the central condensate to the
mass or virial properties of its halo
\cite{PhysRevLett.113.261302,Chan:2021bja,PhysRevD.104.083022,PhysRevD.106.023009}.
In the scalar literature the central object is usually referred to as a
solitonic core, and its mass is compared with a virial mass or with halo
energetic quantities.  The relation has been interpreted both from the scaling
symmetry and uncertainty-principle arguments of the Schr\"odinger-Poisson
system \cite{PhysRevLett.113.261302}, and from the mass gain of cores along merger
histories \cite{Du:2016aik}.  It therefore provides a compact way to
characterize how self-gravitating wave cores are
embedded in their
surrounding halos, and has been applied in many different ways to understand the phenomenology of solitons in astrophysics e.g. Refs.~\cite{Hui:2016ltb,Bar:2018acw,Bar:2019pnz,Du:2023jxh,Escudero:2023vgv,Teodori:2025rul}.  However, it is important to note that simulations indicate a large degree of scatter in the core-halo relation away from analytic estimates~\cite{Chan:2021bja}, possibly driven by environmental effects and merger history, indicating the need for dedicated cosmological studies in specific models.

Light spin-1 fields provide another important realization of ultralight
bosonic dark matter.  Such particles may arise as hidden or dark photons
\cite{Arias:2012az,Graham:2015rva,Goodsell:2009xc}, and their phenomenology has
attracted increasing attention in both cosmology and direct detection
\cite{Caputo:2021eaa}.  In this work we refer to these models generically as
vector dark matter.  Vector dark matter can also form compact
self-gravitating solitons, usually called Proca stars
\cite{Brito:2015pxa,Herdeiro:2016tmi,Sanchis-Gual:2017bhw,DiGiovanni:2018qxl,Jain:2021pnk}.

For both scalar and vector ultralight dark matter, the large occupation number
allows the field to be treated classically on astrophysical scales.  In the
nonrelativistic regime, the dynamics are described by Schr\"odinger-type
equations coupled to gravity.  The wave nature of the field leads to phenomena
absent in collisionless cold dark matter, including gradient-pressure support,
interference patterns, and the formation of compact self-gravitating
condensates.  In the vector case, the massive spin-1 field has three physical
degrees of freedom and is described by a complex three-component wavefunction.
Gravity couples to the total density of the three components, while their
relative amplitudes and phases encode the polarization and spin structure of
the field.

Simulations have shown that Proca stars can form dynamically in vector
dark matter \cite{Gorghetto:2022sue,PhysRevD.108.083021}.  Related
work has investigated gravitational condensation of vector fields, the
polarization structure of vector solitons, and the dynamics of vector
dark-matter halos
\cite{Zhang:2021xxa,Adshead:2021kvl,PhysRevD.108.083021,PhysRevD.111.043031,Glennon:2023jsp}.

However, the scalar core-halo relation cannot simply be carried over to the
vector case\cite{Amin:2022pzv}.  A Proca condensate is built from three coupled wave components,
and the local polarization and spin structure can evolve during evolution.  It is therefore necessary
to measure directly how Proca-star cores are embedded in cosmological vector
halos.
In this paper, we use three-component Schr\"odinger-Poisson simulations with
cosmological initial conditions~\cite{Gorghetto:2022sue} to identify halos containing compact central
Proca-star condensates and to investigate the 
mass relation between core (Proca star) and halo. We further characterize the evolution of the global polarization fractions and the average spin density of individual halos over cosmic time.

The remainder of this paper is organized as follows.  In
Sec.~\ref{sec:theory}, we introduce the theoretical and numerical framework:
Sec.~\ref{sec:eom} summarizes the nonrelativistic vector
Schr\"odinger--Poisson system and the cosmological initial conditions, while
Sec.~\ref{sec:numerical_method} describes the numerical simulations and the
range over which the results are considered reliable.  Section~\ref{sec:results} presents our results: in
Sec.~\ref{sec:corehalo}, we present the measured core--halo mass relation for
profile-confirmed Proca-star condensates; Sec.~\ref{sec:mass_function} studies the halo mass function and Proca star occupation fraction; Sec.~\ref{sec:polarization_check}
examines the evolution of the global longitudinal and transverse polarization
fractions; and Sec.~\ref{sec:halo_spin} investigates the local spin evolution
within individual halos.  We summarize our main findings and discuss future
directions in Sec.~\ref{sec:conclusions}.  Additional details of the halo and
Proca star identification procedure are provided in
Appendix~\ref{app:catalog}.

\section{Theory and Simulation of Vector Dark Matter}
\label{sec:theory}
\subsection{Equations and initial conditions}
\label{sec:eom}

We consider a real massive vector field \(A_\mu\) minimally coupled to gravity,
\begin{equation}
S
=
\int \mathrm{d}^4x\,\sqrt{-g}
\left[
-\frac14 F_{\mu\nu}F^{\mu\nu}
-\frac12 m^2 A_\mu A^\mu
\right],
\label{eq:proca_action}
\end{equation}
where
\begin{equation}
F_{\mu\nu}
=
\partial_\mu A_\nu-\partial_\nu A_\mu .
\end{equation}
The corresponding equation of motion is
\begin{equation}
\nabla_\mu F^{\mu\nu}
=
m^2 A^\nu ,
\label{eq:proca_eq}
\end{equation}
together with the Proca constraint \(\nabla_\mu A^\mu=0\), which leaves three
physical degrees of freedom for \(m\neq0\).

{
In the nonrelativistic regime these degrees of freedom may be written in terms
of slowly varying complex amplitudes.  For the spatial components we use
\begin{align}
A_i
&=
\frac{1}{\sqrt{2m^2a^3}}
\left(\psi_i e^{-imt}+\psi_i^*e^{imt}\right),
\qquad i=1,2,3 .
\label{eq:proca_to_wavefunction}
\end{align}
The factor \(a^{-3/2}\) removes the dilution associated with the homogeneous
expansion.  To leading order in the weak-field and nonrelativistic expansion,
the complex vector wavefunction
\begin{equation}
\bm{\psi}
=
(\psi_x,\psi_y,\psi_z)^T
\end{equation}
obeys
\begin{align}
i\frac{\partial\bm{\psi}}{\partial t}
&=
-\frac{1}{2ma^2}\nabla^2\bm{\psi}
+
m\Phi\,\bm{\psi} ,
\label{eq:sp_physical_1}
\\
\nabla^2\Phi
&=
\frac{4\pi G}{a}
\left[
\bm{\psi}^\dagger\bm{\psi}
-
\left\langle\bm{\psi}^\dagger\bm{\psi}\right\rangle
\right] .
\label{eq:sp_physical_2}
\end{align}
Here \(\bm{x}\) is the comoving coordinate, \(\nabla\) denotes the comoving
derivative, and \(\Phi\) is the Newtonian potential.  The density sourcing the
gravitational potential is
\begin{equation}
\rho_{\rm phys}(\bm{x})
=
\frac{\bm{\psi}^\dagger\bm{\psi}}{a^3}
=
\frac{1}{a^3}\sum_{i=x,y,z}|\psi_i|^2 .
\label{eq:physical_density}
\end{equation}
Thus the gravitational potential depends only on the total density, whereas
the vector nature of the field is contained in the relative phases and
amplitudes of the three components.
}

{
For the homogeneous expansion we include matter and radiation.  The Friedmann equation used to
compute the background scale factor is
\begin{equation}
H^2(a)
=
\frac{H_{\rm eq}^2}{2}
\left(a^{-3}+a^{-4}\right)
=
\frac{H_{\rm eq}^2}{2a^4}(1+a),
\label{eq:friedmann_mr}
\end{equation}
where \(H_{\rm eq}\) is the Hubble rate at equality.  With this convention
\(a=1\) corresponds to matter-radiation equality, while \(a<1\) and
\(a>1\) denote the radiation- and matter-dominated sides of the transition.
For this background the scale factor is evaluated analytically using the
stored background time variable \(u\),
\begin{equation}
a(u)=\frac12 u(1+u).
\label{eq:scale_factor_code}
\end{equation}
The production runs start at \(a=0.01\), well before equality; the outputs used for
the core-halo and spin analysis are after equality.
}

{
We introduce the dimensionless variables
\begin{align}
\widetilde t
&=
\int \frac{\mathrm{d} t}{T_{\rm norm}a^2},
&
\widetilde{\bm{x}}
&=
\frac{\bm{x}}{\sqrt{T_{\rm norm}/m}},
\nonumber\\
\widetilde{\Phi}
&=
mT_{\rm norm}a^2\Phi,
&
\widetilde{\bm{\psi}}
&=
\sqrt{4\pi G}\,T_{\rm norm}\bm{\psi} .
\label{eq:dimensionless_variables}
\end{align}
Taking \(T_{\rm norm}=H_{\rm eq}^{-1}\),
Eqs.~\eqref{eq:sp_physical_1} and \eqref{eq:sp_physical_2} become
\begin{align}
i\frac{\partial\widetilde{\bm{\psi}}}{\partial\widetilde t}
&=
-\frac12\widetilde\nabla^2\widetilde{\bm{\psi}}
+
\widetilde\Phi\,\widetilde{\bm{\psi}} ,
\label{eq:dimensionless_sp_1}
\\
\widetilde\nabla^2\widetilde\Phi
&=
a\left(
|\widetilde{\bm{\psi}}|^2
-\left\langle|\widetilde{\bm{\psi}}|^2\right\rangle
\right).
\label{eq:dimensionless_sp_2}
\end{align}
Unless otherwise stated, all quantities below are quoted in
these dimensionless variables.
}

The initial conditions are generated using the same class of cosmological
vector-dark-matter initial conditions as in Ref.~\cite{Gorghetto:2022sue}.
The resulting three-component wavefunction is evolved with
Eqs.~\eqref{eq:dimensionless_sp_1} and \eqref{eq:dimensionless_sp_2}.  The
normalization is fixed by setting the vector dark-matter fraction to
\(f=\Omega_A/\Omega_M=0.84\).  The remaining matter component contributes to
the homogeneous expansion but is not assigned independent inhomogeneous
degrees of freedom in these simulations.  For reference, the input spectrum
is characterized by
\begin{equation}
P_{A_L}^{\rm IC}(k)
\propto
\frac{2\pi^2}{k^3}
\frac{(k/k_\star)^2}{1+(k/k_\star)^3},
\label{eq:initial_power_spectrum}
\end{equation}
where \(k_\star\) is the characteristic peak scale fixed by the
initial-condition construction.  

The spectrum in
Eq.~\eqref{eq:initial_power_spectrum} is derived from isocurvature fluctuations
in the vector field generated during inflation and peaks at \(k_\star\), in
contrast to the nearly scale-invariant spectrum of a scalar field
\cite{Gorghetto:2022sue}.  This enhancement over ordinary adiabatic curvature
fluctuations leads to an early period of structure formation similar to that
seen with axion miniclusters
\cite{Hogan:1988mp,Kolb:1994fi,Ellis:2020gtq,Eggemeier:2019jsu}.  Vector-dark-
matter structure formation with Eq.~\eqref{eq:initial_power_spectrum} is
similar to axion miniclusters for a temperature-independent axion mass, where
the Jeans scale lies near the peak of the power spectrum
\cite{OHare:2021zrq}, unlike QCD axion miniclusters, for which the
temperature-dependent mass separates these scales. Thus, Proca star formation with these initial conditions is enhanced relative to axion star formation inside QCD axion miniclusters~\cite{Gorghetto:2022sue}.

The
following analysis takes the initial power spectrum defined at matter-radiation equality, where fluctuations are linear, and evolves the vector dark matter into the non-linear regime of halo formation.

\subsection{Numerical simulations}
\label{sec:numerical_method}
We solve Eqs.~\eqref{eq:dimensionless_sp_1} and
\eqref{eq:dimensionless_sp_2} with a pseudospectral method based on Ref.~\cite{PhysRevD.108.083021} to follow the nonlinear cosmological evolution
of vector dark matter
\cite{Gorghetto:2022sue,PhysRevD.104.083022,PhysRevD.108.083021,Zeng:2025unb,Chen:2026sph}.
We solve the vector Schr\"odinger--Poisson equations using a fourth-order
pseudospectral method with periodic boundary conditions, considering
dimensionless box sizes \(\widetilde L\in[10,50]\) with $512^3$ and $256^3$ grids.
The time variable used below is the equality-normalized scale factor defined
in Eq.~\eqref{eq:friedmann_mr}.  Using \(z_{\rm eq}=3370\), the corresponding
redshift is
\begin{equation}
1+z=\frac{1+z_{\rm eq}}{a}.
\label{eq:a_to_z}
\end{equation}
Thus the last output used in the quantitative analysis, \(a\simeq4.3\),
corresponds to \(z\simeq7.9\times10^2\), identified as the lowest redshift
where the box scale remains in the linear regime of structure formation.

The dimensionless box length can also be converted to a comoving physical
length after specifying the vector mass.  With the notation of
Ref.~\cite{Gorghetto:2022sue}, \(\lambda_\star=2\pi/k_\star\), and our
dimensionless coordinate corresponds to
\(\ell_0=\lambda_\star/(2\pi)\).  Since
\(a_0\lambda_\star\simeq10^{11}{\rm km}\,(10^{-5}{\rm eV}/m)^{1/2}\), we have
\begin{equation}
L_{\rm com}
=
\widetilde L\,\ell_0
\simeq
5.16\times10^{-10}\,\widetilde L
\left(\frac{10^{-5}{\rm eV}}{m}\right)^{1/2}
{\rm Mpc}.
\label{eq:L_to_mpc}
\end{equation}
For example, the boxes with \(\widetilde L=26\)--\(40\) correspond to
\(L_{\rm com}\simeq(1.34\text{--}2.06)\times10^{-8}
\left(10^{-5}{\rm eV}/m\right)^{1/2}{\rm Mpc}\).

\begin{figure}[tbp]
\centering
\safeincludegraphics[width=0.92\columnwidth]{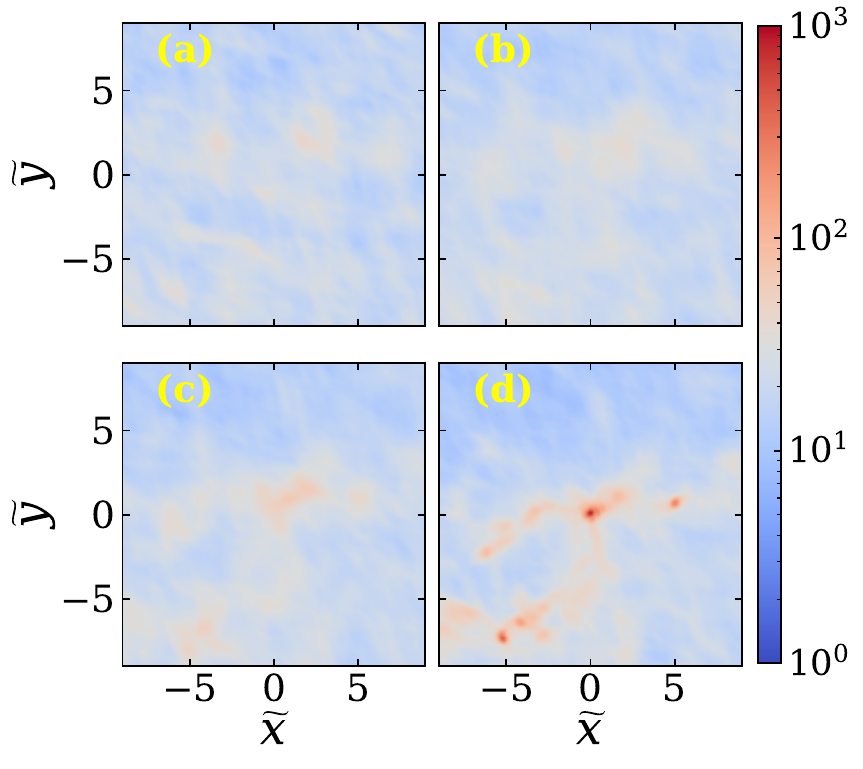}
\caption{
Snapshots of the projected density field from the $\widetilde{L}=30$
simulation.
(a) $a=0.01$ ($z\simeq3.4\times10^5$).
(b) $a\simeq2.15$ ($z\simeq1.6\times10^3$).
(c) $a\simeq3.24$ ($z\simeq1.0\times10^3$).
(d) $a\simeq4.3$ ($z\simeq7.8\times10^2$).
}
\label{fig:density_projection}
\end{figure}

The simulations are evolved beyond this time, but the final outputs are
affected by finite-volume and resolution limitations.  {As a
consistency check, we verified that, on the largest scales, all production
boxes are converged and follow the expected linear growth through
the last trusted snapshot, \(a\simeq4.3\).  We therefore restrict all
quantitative measurements below to outputs with \(a\lesssim4.3\).}  One example is shown in
Fig.~\ref{fig:density_projection},
which shows the projected density in the \(\widetilde{L}=30\) simulation at
four representative scale factors within this trusted interval.  Nonlinear
overdense regions form and develop compact central condensates embedded in
extended halos and the filamentary cosmic web at later times, see Fig.~\ref{fig:density_projection}(c) and (d).

After the compact condensates have formed, we further check their inner
structure by fitting the spherically averaged density profiles around the
halo centers.  For representative halos in the \(\widetilde L=30\)
simulation at \(a\simeq4.3\), the central profiles are well described by the
same soliton-like form used in Appendix~\ref{app:catalog}, as illustrated in
Fig.~\ref{fig:soliton_profile_fit}.  

\begin{figure}[tbp]
\centering
\safeincludegraphics[width=0.94\columnwidth]{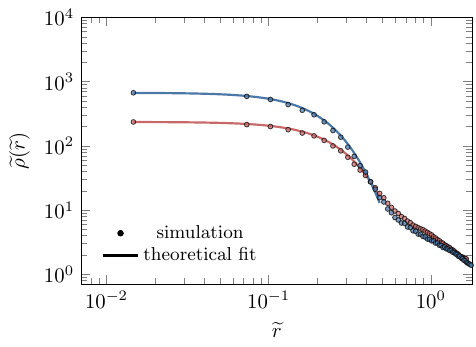}
\caption{
Representative radial density profiles for two halos in the
\(\widetilde L=30\) simulation at
\(a\simeq4.3\).  The red and blue profiles correspond to
\((\widetilde M_{\rm h},\widetilde M_\star)=(40.9,26.1)\) and
\((57.9,42.1)\), respectively.  Points show the spherically averaged
simulation profiles, and solid curves show the theoretical profile fits.
For each curve, both the central density normalization and the soliton
radius are fitted to the inner profile. 
}
\label{fig:soliton_profile_fit}
\end{figure}

\section{Results}
\label{sec:results}
\subsection{Core-halo relation}
\label{sec:corehalo}

At \(a\simeq4.3\), the combined
profile-confirmed sample contains \(\sim300\) objects.  To quote a single
core-halo law, we calibrate the baseline relation at this output, where the
central condensates are most clearly developed, and fit
\begin{equation}
\widetilde{M}_\star
=
A\,\widetilde{M}_{\rm h}^\alpha .
\label{eq:corehalo_model}
\end{equation}
The reported coefficients are obtained by ordinary least squares in
logarithmic variables using the combined profile-confirmed sample at the
final analyzed output.

We now present the relation between the mass of the Proca star and the
mass of its halo.  The measured soliton-halo pairs at \(a\simeq4.3\) are fitted
with Eq.~\eqref{eq:corehalo_model}, and the best-fit parameters are listed in
Table~\ref{tab:corehalo_params}.

\begin{figure*}[tbp]
\centering
\safeincludegraphics[width=0.88\textwidth]{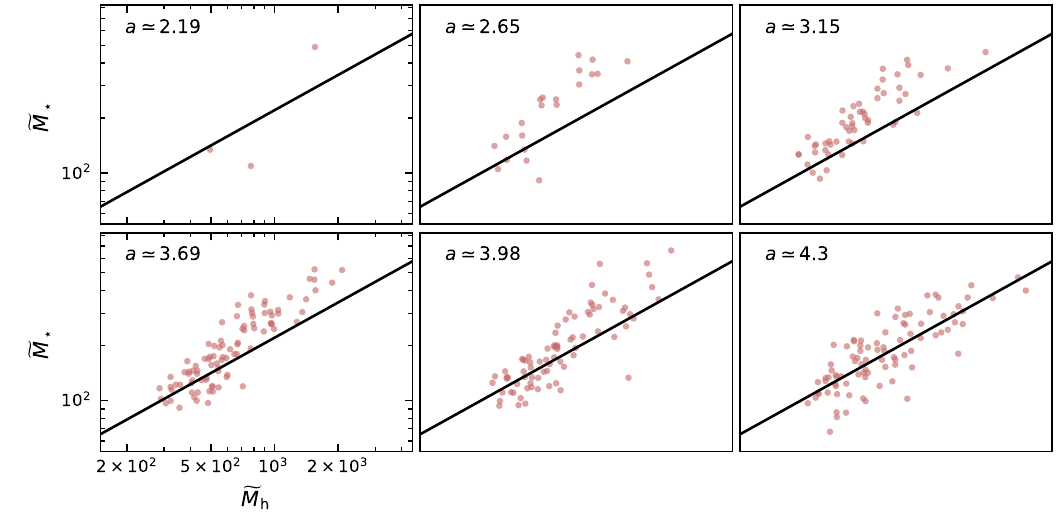}
\caption{
Core-halo measurements in the \(\widetilde{L}=30\) simulation at different
scale factors.  Points denote profile-confirmed soliton-halo mass pairs, while
solid curves show Eq.~\eqref{eq:corehalo_model}.
}
\label{fig:corehalo_fit}
\end{figure*}

\begin{figure}[tbp]
\centering
\safeincludegraphics[width=0.92\columnwidth]{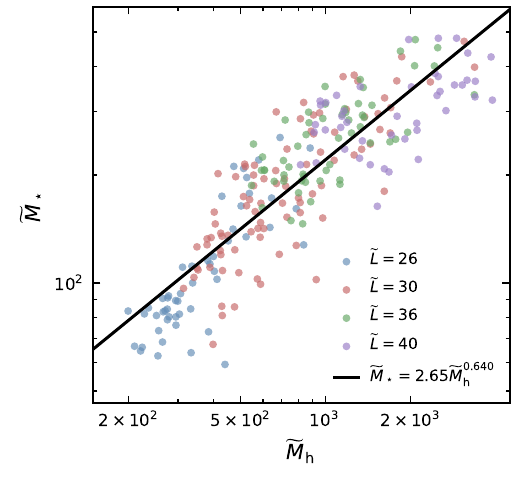}
\caption{
Combined core-halo measurements from the
simulations with different $\widetilde{L}$ at $a\simeq4.3$.  Points show
profile-confirmed soliton-halo pairs, and the solid curve shows
Eq.~\eqref{eq:corehalo_model}.
}
\label{fig:corehalo_all_boxes}
\end{figure}

Fig.~\ref{fig:corehalo_fit} shows representative measurements at six scale
factors in the \(\widetilde{L}=30\) simulation.  At earlier times, fewer
profile-confirmed objects are identified.  At later times, the sample becomes
more populated, and the confirmed soliton masses show a positive correlation
with host-halo mass.  Fig.~\ref{fig:corehalo_all_boxes} shows the corresponding
measurements from simulations with different box sizes at \(a\simeq4.3\)
(\(z\simeq7.9\times10^2\)).  We find that
Eq.~\eqref{eq:corehalo_model} provides a good fit to these data. The profile-integrated masses give
\(\alpha=0.6403\pm0.0235\), with a logarithmic scatter of \(0.109\) dex.

\begin{table}[!b]
\centering
\begin{tabular}{ccc}
\toprule
Parameter & Best-fit value & Interpretation \\
\midrule
$A$      & $2.65\pm0.42$  & Normalization \\
$\alpha$ & $0.6403\pm0.0235$   & Halo-mass dependence \\
\bottomrule
\end{tabular}
\caption{
Best-fit parameters and $1\sigma$ uncertainties for
Eq.~\eqref{eq:corehalo_model}, using the profile-confirmed sample at
\(a\simeq4.3\).
}
\label{tab:corehalo_params}
\end{table}

\subsection{Halo mass function and Proca-star occupation}
\label{sec:mass_function}

\begin{figure}[tbp]
\centering
\safeincludegraphics[width=0.94\columnwidth]{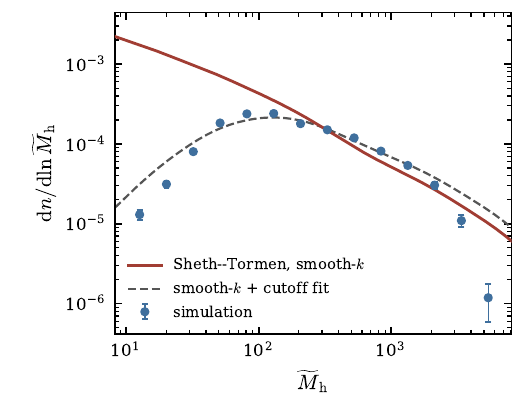}
\caption{
Halo mass function at \(a\simeq4.3\).  Points show the simulation catalogue,
the solid curve shows the smooth-\(k\) Sheth--Tormen result, and the dashed
curve includes the low-mass suppression in
Eq.~\eqref{eq:hmf_low_mass_cutoff}.
}
\label{fig:halo_mass_function}
\end{figure}

The full halo catalogue gives the halo mass function at \(a\simeq4.3\), shown
in Fig.~\ref{fig:halo_mass_function}.  We compute the variance from the
linear density power spectrum as
\begin{equation}
\sigma^2(R)=
\int_0^\infty\frac{k^2\,\mathrm{d}k}{2\pi^2}
P_{\rm lin}(k)\,\widetilde W^2(kR),
\label{eq:hmf_variance}
\end{equation}
where the spectrum measured at \(a=0.01\) is evolved to the epoch of the
catalogue using the linear growth factor \(D(a)=1+3a/2\)
\cite{Ellis:2020gtq}.  Computing the theoretical mass function in models with a truncated initial spectrum is best done using a
smooth-\(k\) filter~\cite{Du:2023jxh},
\begin{equation}
\widetilde W_{\rm sk}(kR)=\frac{1}{1+(kR)^\beta},
\qquad
\widetilde M=\frac{4\pi}{3}(c_W R)^3\overline{\widetilde\rho},
\label{eq:smooth_k_window}
\end{equation}
with \(\beta=9.10\) and \(c_W=2.16\).  The corresponding Sheth--Tormen
mass function is \cite{Sheth:1999mn}
\begin{align}
\frac{\mathrm{d}n}{\mathrm{d}\ln\widetilde M}
&=\frac{\overline{\widetilde\rho}}{\widetilde M}
f_{\rm ST}(\sigma)
\left|\frac{\mathrm{d}\ln\sigma^{-1}}{\mathrm{d}\ln\widetilde M}\right|,
\nonumber\\
f_{\rm ST}(\sigma)
&=A_{\rm ST}\sqrt{\frac{2q}{\pi}}\frac{\delta_c}{\sigma}
\left[1+\left(\frac{\sigma^2}{q\delta_c^2}\right)^p\right]
\exp\left(-\frac{q\delta_c^2}{2\sigma^2}\right),
\label{eq:sheth_tormen_hmf}
\end{align}
where \((A_{\rm ST},q,p)=(0.3222,0.707,0.3)\).  We use the
era collapse threshold
\(\delta_c(a)=1.686D(a)/[D(a)-1]\simeq1.95\), which is valid across the radiation-matter transition~\cite{Ellis:2020gtq}.
The solid curve in Fig.~\ref{fig:halo_mass_function} uses this prescription,
with a normalization factor of \(1.09\) determined from
\(\widetilde M_{\rm h}\geq200\). 

We observe that even using the smooth-$k$ window function, our computation does not fit the simulated mass function, which has a turnover at low mass not predicted by the theoretical model. The low mass turnover emerges directly from our field-based halo finder, and due to the de Broglie wavelength (Jeans) smoothing inherent in the Schr\"{o}dinger equations, does not suffer from any issues of ``spurious halos'' at low mass (e.g. Ref.~\cite{2016ApJ...818...89S}). We thus fit the residual low-mass turnover with the empirical form~\cite{2016ApJ...818...89S}:
\begin{equation}
\left.\frac{\mathrm{d}n}{\mathrm{d}\ln\widetilde M}\right|_{\rm cut}
=C\left[1+\left(\frac{\widetilde M}{\widetilde M_0}\right)^{-1.1}\right]^{-2.2}
\left.\frac{\mathrm{d}n}{\mathrm{d}\ln\widetilde M}\right|_{\rm ST,sk}.
\label{eq:hmf_low_mass_cutoff}
\end{equation}
A Poisson-likelihood fit gives \(C=1.62\) and
\(\widetilde M_0\simeq69\).  The smooth-\(k\) Sheth--Tormen result describes
the high-mass trend, while Eq.~\eqref{eq:hmf_low_mass_cutoff} provides a
useful description of the additional low-mass suppression in the catalogue. The discrepancy between the simulation and the theory at high halo masses is explained by the finite box size, which does not capture the cosmic population on larges scales.

\begin{figure}[tbp]
\centering
\safeincludegraphics[width=0.94\columnwidth]{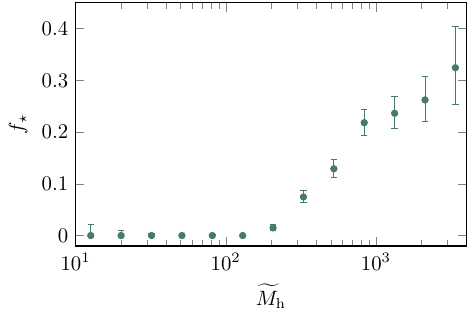}
\caption{
Proca-star occupation fraction \(f_\star\), defined in
Eq.~\eqref{eq:proca_star_fraction}, at \(a\simeq4.3\).  Error bars show
one-standard-deviation binomial Wilson intervals.
}
\label{fig:proca_star_fraction}
\end{figure}

For each halo-mass bin we also define the Proca-star occupation fraction
\begin{equation}
f_\star(\widetilde M_{\rm h})
=\frac{N_\star(\widetilde M_{\rm h})}{N_{\rm h}(\widetilde M_{\rm h})},
\label{eq:proca_star_fraction}
\end{equation}
where \(N_\star\) counts halos satisfying the profile criteria of
Appendix~\ref{app:catalog} and \(N_{\rm h}\) counts all halos in the bin.
The measured fraction is shown separately in Fig.~\ref{fig:proca_star_fraction}.
Interestingly, this fraction remains below \(50\%\) in every mass bin and rises from nearly zero for
\(\widetilde M_{\rm h}\lesssim2\times10^2\) to about \(0.2\)--\(0.3\) in the
well-populated high-mass bins.  Thus the unconfirmed objects are concentrated
toward lower halo masses, although failure of the profile criterion does not
establish the physical absence of a Proca star.

\subsection{Global polarization fractions}
\label{sec:polarization_check}

As a check of the vector-field evolution, we compute the global longitudinal
and transverse polarization fractions of the wavefunction. 

For each snapshot, the Fourier-space wavefunction is decomposed into
components parallel and perpendicular to \(\bm{k}\),
\begin{align}
\widetilde{\bm{\psi}}_L(\bm{k})
&=
\hat{\bm{k}}\left[\hat{\bm{k}}\cdot\widetilde{\bm{\psi}}(\bm{k})\right],
\\
\widetilde{\bm{\psi}}_T(\bm{k})
&=
\widetilde{\bm{\psi}}(\bm{k})-\widetilde{\bm{\psi}}_L(\bm{k}),
\end{align}
where \(\hat{\bm{k}}=\bm{k}/k\).  The corresponding fractions are
\begin{align}
f_L
&=
\frac{\sum_{\bm{k}}|\widetilde{\bm{\psi}}_L(\bm{k})|^2}
{\sum_{\bm{k}}|\widetilde{\bm{\psi}}(\bm{k})|^2},
&
f_T
&=
\frac{\sum_{\bm{k}}|\widetilde{\bm{\psi}}_T(\bm{k})|^2}
{\sum_{\bm{k}}|\widetilde{\bm{\psi}}(\bm{k})|^2} .
\label{eq:global_polarization_fractions}
\end{align}
By construction, \(f_L+f_T=1\) up to numerical roundoff.

\begin{figure}[tbp]
\centering
\safeincludegraphics[width=0.86\columnwidth]{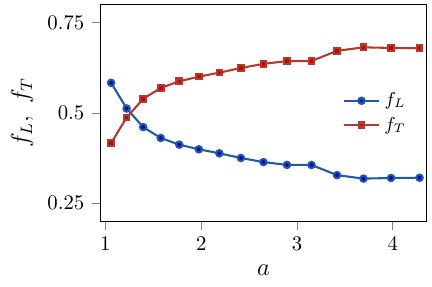}
\caption{
Global Fourier-space polarization fractions, $f_L$ and $f_T$, for the
$\widetilde{L}=30$ simulation with \(a\lesssim4.3\).
}
\label{fig:polarization_time}
\end{figure}

Fig.~\ref{fig:polarization_time} shows the evolution of \(f_L\) and \(f_T\)
in the \(\widetilde{L}=30\) simulation over the trusted interval of scale factor.  The
longitudinal fraction decreases from \(f_L\simeq0.58\) at \(a\simeq1.06\) to
\(f_L\simeq0.32\) at \(a\simeq4.3\), while the transverse fraction increases
accordingly.  The approach toward \(f_L=1/3\) and \(f_T=2/3\) in our simulation
agrees with the previous demonstration in three-dimensional
Schr\"odinger--Poisson simulations~\cite{Amaral_2024}, and is also
qualitatively consistent with Ref.~\cite{Gorghetto:2022sue}.

\subsection{Spin evolution in individual halos}
\label{sec:halo_spin}

The global polarization fractions characterize the Fourier-space content of
the full simulation box.  We now investigate the local spin density,
\begin{equation}
\widetilde{\bm{s}}(\widetilde{\bm{x}})
=
-i\,\widetilde{\bm{\psi}}^*(\widetilde{\bm{x}})\times\widetilde{\bm{\psi}}(\widetilde{\bm{x}}),
\label{eq:spin_density}
\end{equation}
This quantity is evaluated directly from the three complex components of the
wavefunction and measures the local polarization of the vector field, which is of phenomenological relevance as discussed in Ref.~\cite{PhysRevD.111.043031} and references therein.

For the shell-spin diagnostic we use only halos whose central condensates are
profile-confirmed by the soliton fit described in
Appendix~\ref{app:catalog}.  We normalize the radial shells by the same halo
radius \(\widetilde R_{\rm h}\), defined through
\(\delta_{\rm sph}(\widetilde R_{\rm h})=25\), that is used for the halo-mass
measurement.  The halo centers are tracked through the simulation by
recentering on the local density maximum near the previous position.  For
each object, the reference radius measured at \(a\simeq4.3\) is held fixed
when defining the shells at earlier outputs.

For each halo, we measure the density-weighted spin fraction:
\begin{equation}
\left\langle\frac{|\widetilde{\bm{s}}|}{\widetilde{\rho}}\right\rangle_{\widetilde{\rho},{\rm shell}}
=
\frac{\int_{\rm shell} \mathrm{d}^3\widetilde{x}\,|\widetilde{\bm{s}}(\widetilde{\bm{x}})|}
{\int_{\rm shell} \mathrm{d}^3\widetilde{x}\,\widetilde{\rho}(\widetilde{\bm{x}})} .
\label{eq:shell_spin_fraction}
\end{equation}

\begin{figure}[tbp]
\centering
\safeincludegraphics[width=0.94\columnwidth]{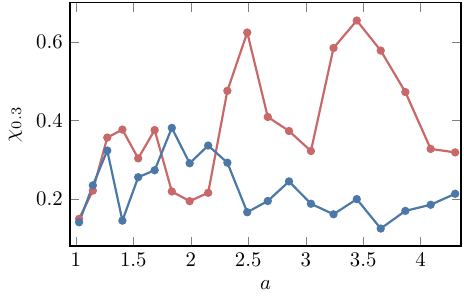}
\caption{
Central spin fraction \(\chi_{0.3}\) for the two halos shown in
Fig.~\ref{fig:soliton_profile_fit}.  The red curve corresponds to
\((\widetilde M_{\rm h},\widetilde M_\star)=(40.9,26.1)\) with
\(\widetilde R_{\rm h}=0.82\), and the blue curve to \((57.9,42.1)\) with
\(\widetilde R_{\rm h}=0.94\).  The colors match Fig.~\ref{fig:soliton_profile_fit}.
}
\label{fig:halo_shell_spin}
\end{figure}

Fig.~\ref{fig:halo_shell_spin} shows examples of the shell-averaged spin
fraction for profile-confirmed halos in the \(\widetilde{L}=30\)
simulation.  For each halo, the average spin fraction in different
radial ranges tends to increase over the trusted interval, although
individual shells show fluctuations.  For the representative halos shown
here, this suggests a gradual build-up of local polarization inside halos
as they form and accrete mass, broadly consistent with the results of Ref.~\cite{PhysRevD.111.043031} for isolated vector dark matter halos.

For the presentation in Fig.~\ref{fig:halo_shell_spin}, we retain only the
cumulative central region, \(\widetilde r<0.3\widetilde R_{\rm h}\), and define
\begin{equation}
\chi_{0.3}
=
\frac{\displaystyle
\int_{\widetilde r<0.3\widetilde R_{\rm h}}
\mathrm{d}^3\widetilde{x}\,|\widetilde{\bm{s}}|}
{\displaystyle
\int_{\widetilde r<0.3\widetilde R_{\rm h}}
\mathrm{d}^3\widetilde{x}\,\widetilde{\rho}} .
\label{eq:central_spin_fraction}
\end{equation}
The two curves correspond to the same profile-confirmed halos shown in
Fig.~\ref{fig:soliton_profile_fit}.

\section{Conclusions}
\label{sec:conclusions}

Using cosmological simulations of the three-component
Schr\"odinger--Poisson equations for vector dark matter, we have studied the
formation of compact Proca-star condensates and their relation to the
surrounding dark-matter halos.  The central objects were identified not only
as local density peaks but also through fits to their spherically averaged
radial density profiles, allowing us to construct a sample of
profile-confirmed cosmological Proca stars.

Our main result is the empirical core--halo mass relation
\[
\widetilde M_\star
=
(2.65\pm0.42)\,
\widetilde M_{\rm h}^{\,0.6403\pm0.0235}
\]
at \(a\simeq4.3\) (redshift $z\simeq 790$), obtained from the combined sample of simulations with
different box sizes.  This result demonstrates a clear positive correlation
between the profile-integrated Proca-star mass and the host-halo mass.  The
fitted exponent is steeper than the
\(M_\star\propto M_{\rm h}^{1/3}\)\cite{PhysRevLett.113.261302} scaling often quoted for scalar wave dark
matter, but consistent within the errors with the maximum slope of \(M_\star\propto M_{\rm h}^{3/5}\) found in the cosmological study of scalar wave dark matter in Ref.~\cite{Chan:2021bja}. Our finding is thus further evidence of the impact cosmology (in terms of initial conditions, halo environment, and halo merger history) as well as the dark matter physics itself, can have on the core-halo mass relation. The cosmological vector initial spectrum, the limited halo-mass
range, and the specific halo definition may also affect the measured
exponent.  We therefore interpret the fitted relation as an empirical result
over the mass and redshift range resolved by the present simulations, rather
than as a universal asymptotic scaling law.

The halo mass function follows the smooth-\(k\) Sheth--Tormen trend at high
masses and exhibits an additional low-mass turnover.  The fraction of halos hosting Proca tars
remains below \(50\%\), although profile confirmation becomes more common with
increasing halo mass.  The unconfirmed objects are concentrated toward lower
masses, although failure of the profile criterion does not demonstrate the
physical absence of a Proca star.


We also followed the global longitudinal and transverse polarization
fractions of the vector field.  Over the analyzed interval, the longitudinal
fraction decreases while the transverse fraction increases, in qualitative
agreement with previous cosmological simulations of vector dark matter~\cite{Amaral_2024,Gorghetto:2022sue}.  In
addition, we examined the shell-averaged spin fraction within
profile-confirmed halos.  For the representative objects studied here, the
spin fraction generally increases during nonlinear evolution, although the
individual radial shells exhibit fluctuations.  These results suggest that
the formation and subsequent evolution of Proca-star halos are accompanied
by a gradual build-up of local polarization and spin structure.

Future simulations with larger volumes, higher spatial resolution, multiple
realizations, and a broader overlapping halo-mass range will be required to
determine whether the fitted core--halo exponent remains stable.  Such
simulations will also allow the redshift evolution of the relation to be
quantified more robustly and enable a closer comparison with halo definitions
commonly adopted in scalar-wave-dark-matter studies.  It will be particularly
interesting to determine whether deviations from the mean relation are
correlated with halo formation history, merger activity, polarization
evolution, or the local spin structure of the central condensate.  These
questions will help establish how the internal vector degrees of freedom
influence the formation and cosmological embedding of Proca stars.

An example of where our result may be of phenomenological interest is in computing the merger rate of Proca stars. Ref.~\cite{Du:2023jxh} found that the slope of the core-halo mass relation is important in determining the merger rate of scalar solitons, which in turn impacts the soliton decay rate via non-gravitational interactions. A steeper slope leads to more rapid mergers, and thus our result suggests that vector dark matter produces a more rapid merger rate of Proca stars than scalar wave dark matter solitons. This could have a number of phenomenological implications in the search for vector dark matter that should be studied further.

\begin{acknowledgments}
We especially thank Zhipan Li for assistance in setting up the computing
clusters, and Marco Gorghetto for helpful correspondence.  J.C. acknowledges support from the Fundamental Research Funds for
the Central Universities under Grant Nos. SWU-KR22012 and SWU-KT25031, and
from the Chongqing Natural Science Foundation General Project under Grant
No. CSTB2023NSCQ-MSX0453.  This work was also facilitated by computational
resources provided by the School of Physical Science and Technology at
Southwest University. D. J. E. M. is supported by an Ernest Rutherford Fellowship (Grant No. ST/T004037/1) and a consolidator grant (Grant No. ST/X000753/1) from the Science and Technologies Facilities Council, United Kingdom. The algorithms and code described here were developed with assistance from Codex, ChatGPT, and Claude Code. We are responsible for the underlying physics, the validation design, and the final review of all content.

\end{acknowledgments}

\bibliographystyle{apsrev4-2}
\bibliography{Ref}

\begin{thebibliography}{56}%
\makeatletter
\providecommand \@ifxundefined [1]{%
 \@ifx{#1\undefined}
}%
\providecommand \@ifnum [1]{%
 \ifnum #1\expandafter \@firstoftwo
 \else \expandafter \@secondoftwo
 \fi
}%
\providecommand \@ifx [1]{%
 \ifx #1\expandafter \@firstoftwo
 \else \expandafter \@secondoftwo
 \fi
}%
\providecommand \natexlab [1]{#1}%
\providecommand \enquote  [1]{``#1''}%
\providecommand \bibnamefont  [1]{#1}%
\providecommand \bibfnamefont [1]{#1}%
\providecommand \citenamefont [1]{#1}%
\providecommand \href@noop [0]{\@secondoftwo}%
\providecommand \href [0]{\begingroup \@sanitize@url \@href}%
\providecommand \@href[1]{\@@startlink{#1}\@@href}%
\providecommand \@@href[1]{\endgroup#1\@@endlink}%
\providecommand \@sanitize@url [0]{\catcode `\\12\catcode `\$12\catcode
  `\&12\catcode `\#12\catcode `\^12\catcode `\_12\catcode `\%12\relax}%
\providecommand \@@startlink[1]{}%
\providecommand \@@endlink[0]{}%
\providecommand \url  [0]{\begingroup\@sanitize@url \@url }%
\providecommand \@url [1]{\endgroup\@href {#1}{\urlprefix }}%
\providecommand \urlprefix  [0]{URL }%
\providecommand \Eprint [0]{\href }%
\providecommand \doibase [0]{https://doi.org/}%
\providecommand \selectlanguage [0]{\@gobble}%
\providecommand \bibinfo  [0]{\@secondoftwo}%
\providecommand \bibfield  [0]{\@secondoftwo}%
\providecommand \translation [1]{[#1]}%
\providecommand \BibitemOpen [0]{}%
\providecommand \bibitemStop [0]{}%
\providecommand \bibitemNoStop [0]{.\EOS\space}%
\providecommand \EOS [0]{\spacefactor3000\relax}%
\providecommand \BibitemShut  [1]{\csname bibitem#1\endcsname}%
\let\auto@bib@innerbib\@empty
\bibitem [{\citenamefont {Aghanim}\ \emph {et~al.}(2020)\citenamefont {Aghanim}
  \emph {et~al.}}]{Aghanim:2018eyx}%
  \BibitemOpen
  \bibfield  {author} {\bibinfo {author} {\bibfnamefont {N.}~\bibnamefont
  {Aghanim}} \emph {et~al.} (\bibinfo {collaboration} {Planck}),\ }\href
  {https://doi.org/10.1051/0004-6361/201833910} {\bibfield  {journal} {\bibinfo
   {journal} {Astron. Astrophys.}\ }\textbf {\bibinfo {volume} {641}},\
  \bibinfo {pages} {A6} (\bibinfo {year} {2020})},\ \Eprint
  {https://arxiv.org/abs/1807.06209} {arXiv:1807.06209 [astro-ph.CO]}
  \BibitemShut {NoStop}%
\bibitem [{\citenamefont {Kimball}\ and\ \citenamefont {van
  Bibber}(2023)}]{Kimball:2023vxk}%
  \BibitemOpen
  \bibinfo {editor} {\bibfnamefont {D.~F.~J.}\ \bibnamefont {Kimball}}\ and\
  \bibinfo {editor} {\bibfnamefont {K.}~\bibnamefont {van Bibber}},\ eds.,\
  \href {https://doi.org/10.1007/978-3-030-95852-7} {\emph {\bibinfo {title}
  {{The Search for Ultralight Bosonic Dark Matter}}}}\ (\bibinfo  {publisher}
  {Springer},\ \bibinfo {year} {2023})\BibitemShut {NoStop}%
\bibitem [{\citenamefont {Marsh}\ \emph {et~al.}(2024)\citenamefont {Marsh},
  \citenamefont {Ellis},\ and\ \citenamefont {Mehta}}]{Marsh:2024ury}%
  \BibitemOpen
  \bibfield  {author} {\bibinfo {author} {\bibfnamefont {D.~J.~E.}\
  \bibnamefont {Marsh}}, \bibinfo {author} {\bibfnamefont {D.}~\bibnamefont
  {Ellis}},\ and\ \bibinfo {author} {\bibfnamefont {V.~M.}\ \bibnamefont
  {Mehta}},\ }\href {https://doi.org/10.1515/9780691249711} {\emph {\bibinfo
  {title} {{Dark Matter: Evidence, Theory, and Constraints}}}},\ Princeton
  Series in Astrophysics\ (\bibinfo  {publisher} {Princeton University Press},\
  \bibinfo {year} {2024})\BibitemShut {NoStop}%
\bibitem [{\citenamefont {Dine}\ and\ \citenamefont
  {Fischler}(1983)}]{Dine:1982ah}%
  \BibitemOpen
  \bibfield  {author} {\bibinfo {author} {\bibfnamefont {M.}~\bibnamefont
  {Dine}}\ and\ \bibinfo {author} {\bibfnamefont {W.}~\bibnamefont
  {Fischler}},\ }\href {https://doi.org/10.1016/0370-2693(83)90639-1}
  {\bibfield  {journal} {\bibinfo  {journal} {Phys. Lett. B}\ }\textbf
  {\bibinfo {volume} {120}},\ \bibinfo {pages} {137} (\bibinfo {year}
  {1983})}\BibitemShut {NoStop}%
\bibitem [{\citenamefont {Preskill}\ \emph {et~al.}(1983)\citenamefont
  {Preskill}, \citenamefont {Wise},\ and\ \citenamefont
  {Wilczek}}]{Preskill:1982cy}%
  \BibitemOpen
  \bibfield  {author} {\bibinfo {author} {\bibfnamefont {J.}~\bibnamefont
  {Preskill}}, \bibinfo {author} {\bibfnamefont {M.~B.}\ \bibnamefont {Wise}},\
  and\ \bibinfo {author} {\bibfnamefont {F.}~\bibnamefont {Wilczek}},\ }\href
  {https://doi.org/10.1016/0370-2693(83)90637-8} {\bibfield  {journal}
  {\bibinfo  {journal} {Phys. Lett. B}\ }\textbf {\bibinfo {volume} {120}},\
  \bibinfo {pages} {127} (\bibinfo {year} {1983})}\BibitemShut {NoStop}%
\bibitem [{\citenamefont {Abbott}\ and\ \citenamefont
  {Sikivie}(1983)}]{Abbott:1982af}%
  \BibitemOpen
  \bibfield  {author} {\bibinfo {author} {\bibfnamefont {L.}~\bibnamefont
  {Abbott}}\ and\ \bibinfo {author} {\bibfnamefont {P.}~\bibnamefont
  {Sikivie}},\ }\href {https://doi.org/10.1016/0370-2693(83)90638-X} {\bibfield
   {journal} {\bibinfo  {journal} {Phys. Lett. B}\ }\textbf {\bibinfo {volume}
  {120}},\ \bibinfo {pages} {133} (\bibinfo {year} {1983})}\BibitemShut
  {NoStop}%
\bibitem [{\citenamefont {Widrow}\ and\ \citenamefont
  {Kaiser}(1993)}]{Widrow:1993qq}%
  \BibitemOpen
  \bibfield  {author} {\bibinfo {author} {\bibfnamefont {L.~M.}\ \bibnamefont
  {Widrow}}\ and\ \bibinfo {author} {\bibfnamefont {N.}~\bibnamefont
  {Kaiser}},\ }\href {https://doi.org/10.1086/187073} {\bibfield  {journal}
  {\bibinfo  {journal} {Astrophys. J. Lett.}\ }\textbf {\bibinfo {volume}
  {416}},\ \bibinfo {pages} {L71} (\bibinfo {year} {1993})}\BibitemShut
  {NoStop}%
\bibitem [{\citenamefont {Suárez}\ \emph {et~al.}(2014)\citenamefont
  {Suárez}, \citenamefont {Robles},\ and\ \citenamefont
  {Matos}}]{Suarez:2013iw}%
  \BibitemOpen
  \bibfield  {author} {\bibinfo {author} {\bibfnamefont {A.}~\bibnamefont
  {Suárez}}, \bibinfo {author} {\bibfnamefont {V.~H.}\ \bibnamefont
  {Robles}},\ and\ \bibinfo {author} {\bibfnamefont {T.}~\bibnamefont
  {Matos}},\ }\href {https://doi.org/10.1007/978-3-319-02063-1\_9} {\bibfield
  {journal} {\bibinfo  {journal} {Astrophys. Space Sci. Proc.}\ }\textbf
  {\bibinfo {volume} {38}},\ \bibinfo {pages} {107} (\bibinfo {year} {2014})},\
  \Eprint {https://arxiv.org/abs/1302.0903} {arXiv:1302.0903 [astro-ph.CO]}
  \BibitemShut {NoStop}%
\bibitem [{\citenamefont {{Guth}}\ \emph {et~al.}(2015)\citenamefont {{Guth}},
  \citenamefont {{Hertzberg}},\ and\ \citenamefont
  {{Prescod-Weinstein}}}]{2015PhRvD..92j3513G}%
  \BibitemOpen
  \bibfield  {author} {\bibinfo {author} {\bibfnamefont {A.~H.}\ \bibnamefont
  {{Guth}}}, \bibinfo {author} {\bibfnamefont {M.~P.}\ \bibnamefont
  {{Hertzberg}}},\ and\ \bibinfo {author} {\bibfnamefont {C.}~\bibnamefont
  {{Prescod-Weinstein}}},\ }\href {https://doi.org/10.1103/PhysRevD.92.103513}
  {\bibfield  {journal} {\bibinfo  {journal} {\prd}\ }\textbf {\bibinfo
  {volume} {92}},\ \bibinfo {eid} {103513} (\bibinfo {year} {2015})},\ \Eprint
  {https://arxiv.org/abs/1412.5930} {arXiv:1412.5930} \BibitemShut {NoStop}%
\bibitem [{\citenamefont {Uhlemann}\ \emph {et~al.}(2014)\citenamefont
  {Uhlemann}, \citenamefont {Kopp},\ and\ \citenamefont
  {Haugg}}]{Uhlemann:2014npa}%
  \BibitemOpen
  \bibfield  {author} {\bibinfo {author} {\bibfnamefont {C.}~\bibnamefont
  {Uhlemann}}, \bibinfo {author} {\bibfnamefont {M.}~\bibnamefont {Kopp}},\
  and\ \bibinfo {author} {\bibfnamefont {T.}~\bibnamefont {Haugg}},\ }\href
  {https://doi.org/10.1103/PhysRevD.90.023517} {\bibfield  {journal} {\bibinfo
  {journal} {\prd}\ }\textbf {\bibinfo {volume} {90}},\ \bibinfo {pages}
  {023517} (\bibinfo {year} {2014})},\ \Eprint
  {https://arxiv.org/abs/1403.5567} {arXiv:1403.5567 [astro-ph.CO]}
  \BibitemShut {NoStop}%
\bibitem [{\citenamefont {Hui}\ \emph {et~al.}(2017)\citenamefont {Hui},
  \citenamefont {Ostriker}, \citenamefont {Tremaine},\ and\ \citenamefont
  {Witten}}]{Hui:2016ltb}%
  \BibitemOpen
  \bibfield  {author} {\bibinfo {author} {\bibfnamefont {L.}~\bibnamefont
  {Hui}}, \bibinfo {author} {\bibfnamefont {J.~P.}\ \bibnamefont {Ostriker}},
  \bibinfo {author} {\bibfnamefont {S.}~\bibnamefont {Tremaine}},\ and\
  \bibinfo {author} {\bibfnamefont {E.}~\bibnamefont {Witten}},\ }\href
  {https://doi.org/10.1103/PhysRevD.95.043541} {\bibfield  {journal} {\bibinfo
  {journal} {Phys. Rev. D}\ }\textbf {\bibinfo {volume} {95}},\ \bibinfo
  {pages} {043541} (\bibinfo {year} {2017})},\ \Eprint
  {https://arxiv.org/abs/1610.08297} {arXiv:1610.08297 [astro-ph.CO]}
  \BibitemShut {NoStop}%
\bibitem [{\citenamefont {Kaup}(1968)}]{Kaup:1968zz}%
  \BibitemOpen
  \bibfield  {author} {\bibinfo {author} {\bibfnamefont {D.~J.}\ \bibnamefont
  {Kaup}},\ }\href {https://doi.org/10.1103/PhysRev.172.1331} {\bibfield
  {journal} {\bibinfo  {journal} {Phys. Rev.}\ }\textbf {\bibinfo {volume}
  {172}},\ \bibinfo {pages} {1331} (\bibinfo {year} {1968})}\BibitemShut
  {NoStop}%
\bibitem [{\citenamefont {Ruffini}\ and\ \citenamefont
  {Bonazzola}(1969)}]{Ruffini:1969qy}%
  \BibitemOpen
  \bibfield  {author} {\bibinfo {author} {\bibfnamefont {R.}~\bibnamefont
  {Ruffini}}\ and\ \bibinfo {author} {\bibfnamefont {S.}~\bibnamefont
  {Bonazzola}},\ }\href {https://doi.org/10.1103/PhysRev.187.1767} {\bibfield
  {journal} {\bibinfo  {journal} {Phys. Rev.}\ }\textbf {\bibinfo {volume}
  {187}},\ \bibinfo {pages} {1767} (\bibinfo {year} {1969})}\BibitemShut
  {NoStop}%
\bibitem [{\citenamefont {Colpi}\ \emph {et~al.}(1986)\citenamefont {Colpi},
  \citenamefont {Shapiro},\ and\ \citenamefont {Wasserman}}]{Colpi:1986ye}%
  \BibitemOpen
  \bibfield  {author} {\bibinfo {author} {\bibfnamefont {M.}~\bibnamefont
  {Colpi}}, \bibinfo {author} {\bibfnamefont {S.~L.}\ \bibnamefont {Shapiro}},\
  and\ \bibinfo {author} {\bibfnamefont {I.}~\bibnamefont {Wasserman}},\ }\href
  {https://doi.org/10.1103/PhysRevLett.57.2485} {\bibfield  {journal} {\bibinfo
   {journal} {Phys. Rev. Lett.}\ }\textbf {\bibinfo {volume} {57}},\ \bibinfo
  {pages} {2485} (\bibinfo {year} {1986})}\BibitemShut {NoStop}%
\bibitem [{\citenamefont {Seidel}\ and\ \citenamefont
  {Suen}(1994)}]{Seidel:1993zk}%
  \BibitemOpen
  \bibfield  {author} {\bibinfo {author} {\bibfnamefont {E.}~\bibnamefont
  {Seidel}}\ and\ \bibinfo {author} {\bibfnamefont {W.-M.}\ \bibnamefont
  {Suen}},\ }\href {https://doi.org/10.1103/PhysRevLett.72.2516} {\bibfield
  {journal} {\bibinfo  {journal} {Phys. Rev. Lett.}\ }\textbf {\bibinfo
  {volume} {72}},\ \bibinfo {pages} {2516} (\bibinfo {year} {1994})},\ \Eprint
  {https://arxiv.org/abs/gr-qc/9309015} {arXiv:gr-qc/9309015} \BibitemShut
  {NoStop}%
\bibitem [{\citenamefont {Liebling}\ and\ \citenamefont
  {Palenzuela}(2012)}]{Liebling:2012fv}%
  \BibitemOpen
  \bibfield  {author} {\bibinfo {author} {\bibfnamefont {S.~L.}\ \bibnamefont
  {Liebling}}\ and\ \bibinfo {author} {\bibfnamefont {C.}~\bibnamefont
  {Palenzuela}},\ }\href {https://doi.org/10.12942/lrr-2012-6} {\bibfield
  {journal} {\bibinfo  {journal} {Living Rev. Rel.}\ }\textbf {\bibinfo
  {volume} {15}},\ \bibinfo {pages} {6} (\bibinfo {year} {2012})},\ \Eprint
  {https://arxiv.org/abs/1202.5809} {arXiv:1202.5809 [gr-qc]} \BibitemShut
  {NoStop}%
\bibitem [{\citenamefont {Schive}\ \emph
  {et~al.}(2014{\natexlab{a}})\citenamefont {Schive}, \citenamefont {Chiueh},\
  and\ \citenamefont {Broadhurst}}]{Schive:2014dra}%
  \BibitemOpen
  \bibfield  {author} {\bibinfo {author} {\bibfnamefont {H.-Y.}\ \bibnamefont
  {Schive}}, \bibinfo {author} {\bibfnamefont {T.}~\bibnamefont {Chiueh}},\
  and\ \bibinfo {author} {\bibfnamefont {T.}~\bibnamefont {Broadhurst}},\
  }\href {https://doi.org/10.1038/nphys2996} {\bibfield  {journal} {\bibinfo
  {journal} {Nature Phys.}\ }\textbf {\bibinfo {volume} {10}},\ \bibinfo
  {pages} {496} (\bibinfo {year} {2014}{\natexlab{a}})},\ \Eprint
  {https://arxiv.org/abs/1406.6586} {arXiv:1406.6586 [astro-ph.GA]}
  \BibitemShut {NoStop}%
\bibitem [{\citenamefont {Schive}\ \emph
  {et~al.}(2014{\natexlab{b}})\citenamefont {Schive}, \citenamefont {Liao},
  \citenamefont {Woo}, \citenamefont {Wong}, \citenamefont {Chiueh},
  \citenamefont {Broadhurst},\ and\ \citenamefont
  {Hwang}}]{PhysRevLett.113.261302}%
  \BibitemOpen
  \bibfield  {author} {\bibinfo {author} {\bibfnamefont {H.-Y.}\ \bibnamefont
  {Schive}}, \bibinfo {author} {\bibfnamefont {M.-H.}\ \bibnamefont {Liao}},
  \bibinfo {author} {\bibfnamefont {T.-P.}\ \bibnamefont {Woo}}, \bibinfo
  {author} {\bibfnamefont {S.-K.}\ \bibnamefont {Wong}}, \bibinfo {author}
  {\bibfnamefont {T.}~\bibnamefont {Chiueh}}, \bibinfo {author} {\bibfnamefont
  {T.}~\bibnamefont {Broadhurst}},\ and\ \bibinfo {author} {\bibfnamefont
  {W.-Y.~P.}\ \bibnamefont {Hwang}},\ }\href
  {https://doi.org/10.1103/PhysRevLett.113.261302} {\bibfield  {journal}
  {\bibinfo  {journal} {Phys. Rev. Lett.}\ }\textbf {\bibinfo {volume} {113}},\
  \bibinfo {pages} {261302} (\bibinfo {year} {2014}{\natexlab{b}})}\BibitemShut
  {NoStop}%
\bibitem [{\citenamefont {Eggemeier}\ and\ \citenamefont
  {Niemeyer}(2019)}]{Eggemeier:2019jsu}%
  \BibitemOpen
  \bibfield  {author} {\bibinfo {author} {\bibfnamefont {B.}~\bibnamefont
  {Eggemeier}}\ and\ \bibinfo {author} {\bibfnamefont {J.~C.}\ \bibnamefont
  {Niemeyer}},\ }\href {https://doi.org/10.1103/PhysRevD.100.063528} {\bibfield
   {journal} {\bibinfo  {journal} {Phys. Rev. D}\ }\textbf {\bibinfo {volume}
  {100}},\ \bibinfo {pages} {063528} (\bibinfo {year} {2019})},\ \Eprint
  {https://arxiv.org/abs/1906.01348} {arXiv:1906.01348 [astro-ph.CO]}
  \BibitemShut {NoStop}%
\bibitem [{\citenamefont {Levkov}\ \emph {et~al.}(2018)\citenamefont {Levkov},
  \citenamefont {Panin},\ and\ \citenamefont {Tkachev}}]{Levkov:2018kau}%
  \BibitemOpen
  \bibfield  {author} {\bibinfo {author} {\bibfnamefont {D.~G.}\ \bibnamefont
  {Levkov}}, \bibinfo {author} {\bibfnamefont {A.~G.}\ \bibnamefont {Panin}},\
  and\ \bibinfo {author} {\bibfnamefont {I.~I.}\ \bibnamefont {Tkachev}},\
  }\href {https://doi.org/10.1103/PhysRevLett.121.151301} {\bibfield  {journal}
  {\bibinfo  {journal} {Phys. Rev. Lett.}\ }\textbf {\bibinfo {volume} {121}},\
  \bibinfo {pages} {151301} (\bibinfo {year} {2018})},\ \Eprint
  {https://arxiv.org/abs/1804.05857} {arXiv:1804.05857 [astro-ph.CO]}
  \BibitemShut {NoStop}%
\bibitem [{\citenamefont {Chavanis}(2021)}]{Chavanis:2020upb}%
  \BibitemOpen
  \bibfield  {author} {\bibinfo {author} {\bibfnamefont {P.-H.}\ \bibnamefont
  {Chavanis}},\ }\href {https://doi.org/10.1140/epjp/s13360-021-01617-3}
  {\bibfield  {journal} {\bibinfo  {journal} {Eur. Phys. J. Plus}\ }\textbf
  {\bibinfo {volume} {136}},\ \bibinfo {pages} {703} (\bibinfo {year}
  {2021})},\ \Eprint {https://arxiv.org/abs/2012.12858} {arXiv:2012.12858
  [astro-ph.GA]} \BibitemShut {NoStop}%
\bibitem [{\citenamefont {Jain}\ \emph {et~al.}(2023)\citenamefont {Jain},
  \citenamefont {Amin}, \citenamefont {Thomas},\ and\ \citenamefont
  {Wanichwecharungruang}}]{Jain:2023ojg}%
  \BibitemOpen
  \bibfield  {author} {\bibinfo {author} {\bibfnamefont {M.}~\bibnamefont
  {Jain}}, \bibinfo {author} {\bibfnamefont {M.~A.}\ \bibnamefont {Amin}},
  \bibinfo {author} {\bibfnamefont {J.}~\bibnamefont {Thomas}},\ and\ \bibinfo
  {author} {\bibfnamefont {W.}~\bibnamefont {Wanichwecharungruang}},\
  }\href@noop {} {\  (\bibinfo {year} {2023})},\ \Eprint
  {https://arxiv.org/abs/2304.01985} {arXiv:2304.01985 [astro-ph.CO]}
  \BibitemShut {NoStop}%
\bibitem [{\citenamefont {Chan}\ \emph {et~al.}(2022)\citenamefont {Chan},
  \citenamefont {Ferreira}, \citenamefont {May}, \citenamefont {Hayashi},\ and\
  \citenamefont {Chiba}}]{Chan:2021bja}%
  \BibitemOpen
  \bibfield  {author} {\bibinfo {author} {\bibfnamefont {H.~Y.~J.}\
  \bibnamefont {Chan}}, \bibinfo {author} {\bibfnamefont {E.~G.~M.}\
  \bibnamefont {Ferreira}}, \bibinfo {author} {\bibfnamefont {S.}~\bibnamefont
  {May}}, \bibinfo {author} {\bibfnamefont {K.}~\bibnamefont {Hayashi}},\ and\
  \bibinfo {author} {\bibfnamefont {M.}~\bibnamefont {Chiba}},\ }\href
  {https://doi.org/10.1093/mnras/stac063} {\bibfield  {journal} {\bibinfo
  {journal} {Mon. Not. Roy. Astron. Soc.}\ }\textbf {\bibinfo {volume} {511}},\
  \bibinfo {pages} {943} (\bibinfo {year} {2022})},\ \Eprint
  {https://arxiv.org/abs/2110.11882} {arXiv:2110.11882 [astro-ph.CO]}
  \BibitemShut {NoStop}%
\bibitem [{\citenamefont {Chen}\ \emph {et~al.}(2021)\citenamefont {Chen},
  \citenamefont {Du}, \citenamefont {Lentz}, \citenamefont {Marsh},\ and\
  \citenamefont {Niemeyer}}]{PhysRevD.104.083022}%
  \BibitemOpen
  \bibfield  {author} {\bibinfo {author} {\bibfnamefont {J.}~\bibnamefont
  {Chen}}, \bibinfo {author} {\bibfnamefont {X.}~\bibnamefont {Du}}, \bibinfo
  {author} {\bibfnamefont {E.~W.}\ \bibnamefont {Lentz}}, \bibinfo {author}
  {\bibfnamefont {D.~J.~E.}\ \bibnamefont {Marsh}},\ and\ \bibinfo {author}
  {\bibfnamefont {J.~C.}\ \bibnamefont {Niemeyer}},\ }\href
  {https://doi.org/10.1103/PhysRevD.104.083022} {\bibfield  {journal} {\bibinfo
   {journal} {Phys. Rev. D}\ }\textbf {\bibinfo {volume} {104}},\ \bibinfo
  {pages} {083022} (\bibinfo {year} {2021})}\BibitemShut {NoStop}%
\bibitem [{\citenamefont {Chen}\ \emph {et~al.}(2022)\citenamefont {Chen},
  \citenamefont {Du}, \citenamefont {Lentz},\ and\ \citenamefont
  {Marsh}}]{PhysRevD.106.023009}%
  \BibitemOpen
  \bibfield  {author} {\bibinfo {author} {\bibfnamefont {J.}~\bibnamefont
  {Chen}}, \bibinfo {author} {\bibfnamefont {X.}~\bibnamefont {Du}}, \bibinfo
  {author} {\bibfnamefont {E.~W.}\ \bibnamefont {Lentz}},\ and\ \bibinfo
  {author} {\bibfnamefont {D.~J.~E.}\ \bibnamefont {Marsh}},\ }\href
  {https://doi.org/10.1103/PhysRevD.106.023009} {\bibfield  {journal} {\bibinfo
   {journal} {Phys. Rev. D}\ }\textbf {\bibinfo {volume} {106}},\ \bibinfo
  {pages} {023009} (\bibinfo {year} {2022})}\BibitemShut {NoStop}%
\bibitem [{\citenamefont {Du}\ \emph {et~al.}(2017)\citenamefont {Du},
  \citenamefont {Behrens}, \citenamefont {Niemeyer},\ and\ \citenamefont
  {Schwabe}}]{Du:2016aik}%
  \BibitemOpen
  \bibfield  {author} {\bibinfo {author} {\bibfnamefont {X.}~\bibnamefont
  {Du}}, \bibinfo {author} {\bibfnamefont {C.}~\bibnamefont {Behrens}},
  \bibinfo {author} {\bibfnamefont {J.~C.}\ \bibnamefont {Niemeyer}},\ and\
  \bibinfo {author} {\bibfnamefont {B.}~\bibnamefont {Schwabe}},\ }\href
  {https://doi.org/10.1103/PhysRevD.95.043519} {\bibfield  {journal} {\bibinfo
  {journal} {Phys. Rev. D}\ }\textbf {\bibinfo {volume} {95}},\ \bibinfo
  {pages} {043519} (\bibinfo {year} {2017})},\ \Eprint
  {https://arxiv.org/abs/1609.09414} {arXiv:1609.09414 [astro-ph.GA]}
  \BibitemShut {NoStop}%
\bibitem [{\citenamefont {Bar}\ \emph {et~al.}(2018)\citenamefont {Bar},
  \citenamefont {Blas}, \citenamefont {Blum},\ and\ \citenamefont
  {Sibiryakov}}]{Bar:2018acw}%
  \BibitemOpen
  \bibfield  {author} {\bibinfo {author} {\bibfnamefont {N.}~\bibnamefont
  {Bar}}, \bibinfo {author} {\bibfnamefont {D.}~\bibnamefont {Blas}}, \bibinfo
  {author} {\bibfnamefont {K.}~\bibnamefont {Blum}},\ and\ \bibinfo {author}
  {\bibfnamefont {S.}~\bibnamefont {Sibiryakov}},\ }\href
  {https://doi.org/10.1103/PhysRevD.98.083027} {\bibfield  {journal} {\bibinfo
  {journal} {Phys. Rev. D}\ }\textbf {\bibinfo {volume} {98}},\ \bibinfo
  {pages} {083027} (\bibinfo {year} {2018})},\ \Eprint
  {https://arxiv.org/abs/1805.00122} {arXiv:1805.00122 [astro-ph.CO]}
  \BibitemShut {NoStop}%
\bibitem [{\citenamefont {Bar}\ \emph {et~al.}(2019)\citenamefont {Bar},
  \citenamefont {Blum}, \citenamefont {Lacroix},\ and\ \citenamefont
  {Panci}}]{Bar:2019pnz}%
  \BibitemOpen
  \bibfield  {author} {\bibinfo {author} {\bibfnamefont {N.}~\bibnamefont
  {Bar}}, \bibinfo {author} {\bibfnamefont {K.}~\bibnamefont {Blum}}, \bibinfo
  {author} {\bibfnamefont {T.}~\bibnamefont {Lacroix}},\ and\ \bibinfo {author}
  {\bibfnamefont {P.}~\bibnamefont {Panci}},\ }\href
  {https://doi.org/10.1088/1475-7516/2019/07/045} {\bibfield  {journal}
  {\bibinfo  {journal} {JCAP}\ }\textbf {\bibinfo {volume} {2019}}\bibfield
  {number} {\bibinfo  {number} { (07)},\ \bibinfo {pages} {045}},\ }\Eprint
  {https://arxiv.org/abs/1905.11745} {arXiv:1905.11745 [astro-ph.CO]}
  \BibitemShut {NoStop}%
\bibitem [{\citenamefont {Du}\ \emph {et~al.}(2024)\citenamefont {Du},
  \citenamefont {Marsh}, \citenamefont {Escudero}, \citenamefont {Benson},
  \citenamefont {Blas}, \citenamefont {Pooni},\ and\ \citenamefont
  {Fairbairn}}]{Du:2023jxh}%
  \BibitemOpen
  \bibfield  {author} {\bibinfo {author} {\bibfnamefont {X.}~\bibnamefont
  {Du}}, \bibinfo {author} {\bibfnamefont {D.~J.~E.}\ \bibnamefont {Marsh}},
  \bibinfo {author} {\bibfnamefont {M.}~\bibnamefont {Escudero}}, \bibinfo
  {author} {\bibfnamefont {A.}~\bibnamefont {Benson}}, \bibinfo {author}
  {\bibfnamefont {D.}~\bibnamefont {Blas}}, \bibinfo {author} {\bibfnamefont
  {C.~K.}\ \bibnamefont {Pooni}},\ and\ \bibinfo {author} {\bibfnamefont
  {M.}~\bibnamefont {Fairbairn}},\ }\href
  {https://doi.org/10.1103/PhysRevD.109.043019} {\bibfield  {journal} {\bibinfo
   {journal} {Phys. Rev. D}\ }\textbf {\bibinfo {volume} {109}},\ \bibinfo
  {pages} {043019} (\bibinfo {year} {2024})},\ \Eprint
  {https://arxiv.org/abs/2301.09769} {arXiv:2301.09769 [astro-ph.CO]}
  \BibitemShut {NoStop}%
\bibitem [{\citenamefont {Escudero}\ \emph {et~al.}(2024)\citenamefont
  {Escudero}, \citenamefont {Pooni}, \citenamefont {Fairbairn}, \citenamefont
  {Blas}, \citenamefont {Du},\ and\ \citenamefont {Marsh}}]{Escudero:2023vgv}%
  \BibitemOpen
  \bibfield  {author} {\bibinfo {author} {\bibfnamefont {M.}~\bibnamefont
  {Escudero}}, \bibinfo {author} {\bibfnamefont {C.~K.}\ \bibnamefont {Pooni}},
  \bibinfo {author} {\bibfnamefont {M.}~\bibnamefont {Fairbairn}}, \bibinfo
  {author} {\bibfnamefont {D.}~\bibnamefont {Blas}}, \bibinfo {author}
  {\bibfnamefont {X.}~\bibnamefont {Du}},\ and\ \bibinfo {author}
  {\bibfnamefont {D.~J.~E.}\ \bibnamefont {Marsh}},\ }\href
  {https://doi.org/10.1103/PhysRevD.109.043018} {\bibfield  {journal} {\bibinfo
   {journal} {Phys. Rev. D}\ }\textbf {\bibinfo {volume} {109}},\ \bibinfo
  {pages} {043018} (\bibinfo {year} {2024})},\ \Eprint
  {https://arxiv.org/abs/2302.10206} {arXiv:2302.10206 [hep-ph]} \BibitemShut
  {NoStop}%
\bibitem [{\citenamefont {Teodori}\ \emph {et~al.}(2026)\citenamefont
  {Teodori}, \citenamefont {Caputo},\ and\ \citenamefont
  {Blum}}]{Teodori:2025rul}%
  \BibitemOpen
  \bibfield  {author} {\bibinfo {author} {\bibfnamefont {L.}~\bibnamefont
  {Teodori}}, \bibinfo {author} {\bibfnamefont {A.}~\bibnamefont {Caputo}},\
  and\ \bibinfo {author} {\bibfnamefont {K.}~\bibnamefont {Blum}},\ }\href
  {https://doi.org/10.1103/jc6p-rlvh} {\bibfield  {journal} {\bibinfo
  {journal} {Phys. Rev. D}\ }\textbf {\bibinfo {volume} {113}},\ \bibinfo
  {pages} {023055} (\bibinfo {year} {2026})},\ \Eprint
  {https://arxiv.org/abs/2501.07631} {arXiv:2501.07631 [astro-ph.GA]}
  \BibitemShut {NoStop}%
\bibitem [{\citenamefont {Arias}\ \emph {et~al.}(2012)\citenamefont {Arias},
  \citenamefont {Cadamuro}, \citenamefont {Goodsell}, \citenamefont {Jaeckel},
  \citenamefont {Redondo},\ and\ \citenamefont {Ringwald}}]{Arias:2012az}%
  \BibitemOpen
  \bibfield  {author} {\bibinfo {author} {\bibfnamefont {P.}~\bibnamefont
  {Arias}}, \bibinfo {author} {\bibfnamefont {D.}~\bibnamefont {Cadamuro}},
  \bibinfo {author} {\bibfnamefont {M.}~\bibnamefont {Goodsell}}, \bibinfo
  {author} {\bibfnamefont {J.}~\bibnamefont {Jaeckel}}, \bibinfo {author}
  {\bibfnamefont {J.}~\bibnamefont {Redondo}},\ and\ \bibinfo {author}
  {\bibfnamefont {A.}~\bibnamefont {Ringwald}},\ }\href
  {https://doi.org/10.1088/1475-7516/2012/06/013} {\bibfield  {journal}
  {\bibinfo  {journal} {JCAP}\ }\textbf {\bibinfo {volume} {2012}}\bibfield
  {number} {\bibinfo  {number} { (06)},\ \bibinfo {pages} {013}},\ }\Eprint
  {https://arxiv.org/abs/1201.5902} {arXiv:1201.5902 [hep-ph]} \BibitemShut
  {NoStop}%
\bibitem [{\citenamefont {Graham}\ \emph {et~al.}(2016)\citenamefont {Graham},
  \citenamefont {Mardon},\ and\ \citenamefont {Rajendran}}]{Graham:2015rva}%
  \BibitemOpen
  \bibfield  {author} {\bibinfo {author} {\bibfnamefont {P.~W.}\ \bibnamefont
  {Graham}}, \bibinfo {author} {\bibfnamefont {J.}~\bibnamefont {Mardon}},\
  and\ \bibinfo {author} {\bibfnamefont {S.}~\bibnamefont {Rajendran}},\ }\href
  {https://doi.org/10.1103/PhysRevD.93.103520} {\bibfield  {journal} {\bibinfo
  {journal} {Phys. Rev. D}\ }\textbf {\bibinfo {volume} {93}},\ \bibinfo
  {pages} {103520} (\bibinfo {year} {2016})},\ \Eprint
  {https://arxiv.org/abs/1504.02102} {arXiv:1504.02102 [hep-ph]} \BibitemShut
  {NoStop}%
\bibitem [{\citenamefont {Goodsell}\ \emph {et~al.}(2009)\citenamefont
  {Goodsell}, \citenamefont {Jaeckel}, \citenamefont {Redondo},\ and\
  \citenamefont {Ringwald}}]{Goodsell:2009xc}%
  \BibitemOpen
  \bibfield  {author} {\bibinfo {author} {\bibfnamefont {M.}~\bibnamefont
  {Goodsell}}, \bibinfo {author} {\bibfnamefont {J.}~\bibnamefont {Jaeckel}},
  \bibinfo {author} {\bibfnamefont {J.}~\bibnamefont {Redondo}},\ and\ \bibinfo
  {author} {\bibfnamefont {A.}~\bibnamefont {Ringwald}},\ }\href
  {https://doi.org/10.1088/1126-6708/2009/11/027} {\bibfield  {journal}
  {\bibinfo  {journal} {JHEP}\ }\textbf {\bibinfo {volume} {2009}}\bibfield
  {number} {\bibinfo  {number} { (11)},\ \bibinfo {pages} {027}},\ }\Eprint
  {https://arxiv.org/abs/0909.0515} {arXiv:0909.0515 [hep-ph]} \BibitemShut
  {NoStop}%
\bibitem [{\citenamefont {Caputo}\ \emph {et~al.}(2021)\citenamefont {Caputo},
  \citenamefont {Millar}, \citenamefont {O'Hare},\ and\ \citenamefont
  {Vitagliano}}]{Caputo:2021eaa}%
  \BibitemOpen
  \bibfield  {author} {\bibinfo {author} {\bibfnamefont {A.}~\bibnamefont
  {Caputo}}, \bibinfo {author} {\bibfnamefont {A.~J.}\ \bibnamefont {Millar}},
  \bibinfo {author} {\bibfnamefont {C.~A.~J.}\ \bibnamefont {O'Hare}},\ and\
  \bibinfo {author} {\bibfnamefont {E.}~\bibnamefont {Vitagliano}},\ }\href
  {https://doi.org/10.1103/PhysRevD.104.095029} {\bibfield  {journal} {\bibinfo
   {journal} {Phys. Rev. D}\ }\textbf {\bibinfo {volume} {104}},\ \bibinfo
  {pages} {095029} (\bibinfo {year} {2021})},\ \Eprint
  {https://arxiv.org/abs/2105.04565} {arXiv:2105.04565 [hep-ph]} \BibitemShut
  {NoStop}%
\bibitem [{\citenamefont {Brito}\ \emph {et~al.}(2016)\citenamefont {Brito},
  \citenamefont {Cardoso}, \citenamefont {Herdeiro},\ and\ \citenamefont
  {Radu}}]{Brito:2015pxa}%
  \BibitemOpen
  \bibfield  {author} {\bibinfo {author} {\bibfnamefont {R.}~\bibnamefont
  {Brito}}, \bibinfo {author} {\bibfnamefont {V.}~\bibnamefont {Cardoso}},
  \bibinfo {author} {\bibfnamefont {C.~A.~R.}\ \bibnamefont {Herdeiro}},\ and\
  \bibinfo {author} {\bibfnamefont {E.}~\bibnamefont {Radu}},\ }\href
  {https://doi.org/10.1016/j.physletb.2015.10.045} {\bibfield  {journal}
  {\bibinfo  {journal} {Phys. Lett. B}\ }\textbf {\bibinfo {volume} {752}},\
  \bibinfo {pages} {291} (\bibinfo {year} {2016})},\ \Eprint
  {https://arxiv.org/abs/1508.05395} {arXiv:1508.05395 [gr-qc]} \BibitemShut
  {NoStop}%
\bibitem [{\citenamefont {Herdeiro}\ \emph {et~al.}(2017)\citenamefont
  {Herdeiro}, \citenamefont {Pombo},\ and\ \citenamefont
  {Radu}}]{Herdeiro:2016tmi}%
  \BibitemOpen
  \bibfield  {author} {\bibinfo {author} {\bibfnamefont {C.~A.~R.}\
  \bibnamefont {Herdeiro}}, \bibinfo {author} {\bibfnamefont {A.~M.}\
  \bibnamefont {Pombo}},\ and\ \bibinfo {author} {\bibfnamefont
  {E.}~\bibnamefont {Radu}},\ }\href
  {https://doi.org/10.1016/j.physletb.2017.10.036} {\bibfield  {journal}
  {\bibinfo  {journal} {Phys. Lett. B}\ }\textbf {\bibinfo {volume} {773}},\
  \bibinfo {pages} {654} (\bibinfo {year} {2017})},\ \Eprint
  {https://arxiv.org/abs/1708.05674} {arXiv:1708.05674 [gr-qc]} \BibitemShut
  {NoStop}%
\bibitem [{\citenamefont {Sanchis-Gual}\ \emph {et~al.}(2017)\citenamefont
  {Sanchis-Gual}, \citenamefont {Herdeiro}, \citenamefont {Radu}, \citenamefont
  {Degollado},\ and\ \citenamefont {Font}}]{Sanchis-Gual:2017bhw}%
  \BibitemOpen
  \bibfield  {author} {\bibinfo {author} {\bibfnamefont {N.}~\bibnamefont
  {Sanchis-Gual}}, \bibinfo {author} {\bibfnamefont {C.}~\bibnamefont
  {Herdeiro}}, \bibinfo {author} {\bibfnamefont {E.}~\bibnamefont {Radu}},
  \bibinfo {author} {\bibfnamefont {J.~C.}\ \bibnamefont {Degollado}},\ and\
  \bibinfo {author} {\bibfnamefont {J.~A.}\ \bibnamefont {Font}},\ }\href
  {https://doi.org/10.1103/PhysRevD.95.104028} {\bibfield  {journal} {\bibinfo
  {journal} {Phys. Rev. D}\ }\textbf {\bibinfo {volume} {95}},\ \bibinfo
  {pages} {104028} (\bibinfo {year} {2017})},\ \Eprint
  {https://arxiv.org/abs/1702.04532} {arXiv:1702.04532 [gr-qc]} \BibitemShut
  {NoStop}%
\bibitem [{\citenamefont {Di~Giovanni}\ \emph {et~al.}(2018)\citenamefont
  {Di~Giovanni}, \citenamefont {Sanchis-Gual}, \citenamefont {Herdeiro},\ and\
  \citenamefont {Font}}]{DiGiovanni:2018qxl}%
  \BibitemOpen
  \bibfield  {author} {\bibinfo {author} {\bibfnamefont {F.}~\bibnamefont
  {Di~Giovanni}}, \bibinfo {author} {\bibfnamefont {N.}~\bibnamefont
  {Sanchis-Gual}}, \bibinfo {author} {\bibfnamefont {C.~A.~R.}\ \bibnamefont
  {Herdeiro}},\ and\ \bibinfo {author} {\bibfnamefont {J.~A.}\ \bibnamefont
  {Font}},\ }\href {https://doi.org/10.1103/PhysRevD.98.064044} {\bibfield
  {journal} {\bibinfo  {journal} {Phys. Rev. D}\ }\textbf {\bibinfo {volume}
  {98}},\ \bibinfo {pages} {064044} (\bibinfo {year} {2018})},\ \Eprint
  {https://arxiv.org/abs/1803.04802} {arXiv:1803.04802 [gr-qc]} \BibitemShut
  {NoStop}%
\bibitem [{\citenamefont {Jain}\ and\ \citenamefont
  {Amin}(2022)}]{Jain:2021pnk}%
  \BibitemOpen
  \bibfield  {author} {\bibinfo {author} {\bibfnamefont {M.}~\bibnamefont
  {Jain}}\ and\ \bibinfo {author} {\bibfnamefont {M.~A.}\ \bibnamefont
  {Amin}},\ }\href {https://doi.org/10.1103/PhysRevD.105.056019} {\bibfield
  {journal} {\bibinfo  {journal} {Phys. Rev. D}\ }\textbf {\bibinfo {volume}
  {105}},\ \bibinfo {pages} {056019} (\bibinfo {year} {2022})},\ \Eprint
  {https://arxiv.org/abs/2109.04892} {arXiv:2109.04892 [hep-th]} \BibitemShut
  {NoStop}%
\bibitem [{\citenamefont {Gorghetto}\ \emph {et~al.}(2022)\citenamefont
  {Gorghetto}, \citenamefont {Hardy}, \citenamefont {March-Russell},
  \citenamefont {Song},\ and\ \citenamefont {West}}]{Gorghetto:2022sue}%
  \BibitemOpen
  \bibfield  {author} {\bibinfo {author} {\bibfnamefont {M.}~\bibnamefont
  {Gorghetto}}, \bibinfo {author} {\bibfnamefont {E.}~\bibnamefont {Hardy}},
  \bibinfo {author} {\bibfnamefont {J.}~\bibnamefont {March-Russell}}, \bibinfo
  {author} {\bibfnamefont {N.}~\bibnamefont {Song}},\ and\ \bibinfo {author}
  {\bibfnamefont {S.~M.}\ \bibnamefont {West}},\ }\href
  {https://doi.org/10.1088/1475-7516/2022/08/018} {\bibfield  {journal}
  {\bibinfo  {journal} {JCAP}\ }\textbf {\bibinfo {volume} {2022}}\bibfield
  {number} {\bibinfo  {number} { (08)},\ \bibinfo {pages} {018}},\ }\Eprint
  {https://arxiv.org/abs/2203.10100} {arXiv:2203.10100 [hep-ph]} \BibitemShut
  {NoStop}%
\bibitem [{\citenamefont {Chen}\ \emph {et~al.}(2023)\citenamefont {Chen},
  \citenamefont {Du}, \citenamefont {Zhou}, \citenamefont {Benson},\ and\
  \citenamefont {Marsh}}]{PhysRevD.108.083021}%
  \BibitemOpen
  \bibfield  {author} {\bibinfo {author} {\bibfnamefont {J.}~\bibnamefont
  {Chen}}, \bibinfo {author} {\bibfnamefont {X.}~\bibnamefont {Du}}, \bibinfo
  {author} {\bibfnamefont {M.}~\bibnamefont {Zhou}}, \bibinfo {author}
  {\bibfnamefont {A.}~\bibnamefont {Benson}},\ and\ \bibinfo {author}
  {\bibfnamefont {D.~J.~E.}\ \bibnamefont {Marsh}},\ }\href
  {https://doi.org/10.1103/PhysRevD.108.083021} {\bibfield  {journal} {\bibinfo
   {journal} {Phys. Rev. D}\ }\textbf {\bibinfo {volume} {108}},\ \bibinfo
  {pages} {083021} (\bibinfo {year} {2023})}\BibitemShut {NoStop}%
\bibitem [{\citenamefont {Zhang}\ \emph {et~al.}(2022)\citenamefont {Zhang},
  \citenamefont {Jain},\ and\ \citenamefont {Amin}}]{Zhang:2021xxa}%
  \BibitemOpen
  \bibfield  {author} {\bibinfo {author} {\bibfnamefont {H.-Y.}\ \bibnamefont
  {Zhang}}, \bibinfo {author} {\bibfnamefont {M.}~\bibnamefont {Jain}},\ and\
  \bibinfo {author} {\bibfnamefont {M.~A.}\ \bibnamefont {Amin}},\ }\href
  {https://doi.org/10.1103/PhysRevD.105.096037} {\bibfield  {journal} {\bibinfo
   {journal} {Phys. Rev. D}\ }\textbf {\bibinfo {volume} {105}},\ \bibinfo
  {pages} {096037} (\bibinfo {year} {2022})},\ \Eprint
  {https://arxiv.org/abs/2111.08700} {arXiv:2111.08700 [astro-ph.CO]}
  \BibitemShut {NoStop}%
\bibitem [{\citenamefont {Adshead}\ and\ \citenamefont
  {Lozanov}(2021)}]{Adshead:2021kvl}%
  \BibitemOpen
  \bibfield  {author} {\bibinfo {author} {\bibfnamefont {P.}~\bibnamefont
  {Adshead}}\ and\ \bibinfo {author} {\bibfnamefont {K.~D.}\ \bibnamefont
  {Lozanov}},\ }\href {https://doi.org/10.1103/PhysRevD.103.103501} {\bibfield
  {journal} {\bibinfo  {journal} {Phys. Rev. D}\ }\textbf {\bibinfo {volume}
  {103}},\ \bibinfo {pages} {103501} (\bibinfo {year} {2021})},\ \Eprint
  {https://arxiv.org/abs/2101.07265} {arXiv:2101.07265 [gr-qc]} \BibitemShut
  {NoStop}%
\bibitem [{\citenamefont {Chen}\ \emph {et~al.}(2025)\citenamefont {Chen},
  \citenamefont {Nguyen},\ and\ \citenamefont {Marsh}}]{PhysRevD.111.043031}%
  \BibitemOpen
  \bibfield  {author} {\bibinfo {author} {\bibfnamefont {J.}~\bibnamefont
  {Chen}}, \bibinfo {author} {\bibfnamefont {L.~H.}\ \bibnamefont {Nguyen}},\
  and\ \bibinfo {author} {\bibfnamefont {D.~J.~E.}\ \bibnamefont {Marsh}},\
  }\href {https://doi.org/10.1103/PhysRevD.111.043031} {\bibfield  {journal}
  {\bibinfo  {journal} {Phys. Rev. D}\ }\textbf {\bibinfo {volume} {111}},\
  \bibinfo {pages} {043031} (\bibinfo {year} {2025})}\BibitemShut {NoStop}%
\bibitem [{\citenamefont {Glennon}\ \emph {et~al.}(2023)\citenamefont
  {Glennon}, \citenamefont {Musoke},\ and\ \citenamefont
  {Prescod-Weinstein}}]{Glennon:2023jsp}%
  \BibitemOpen
  \bibfield  {author} {\bibinfo {author} {\bibfnamefont {N.}~\bibnamefont
  {Glennon}}, \bibinfo {author} {\bibfnamefont {N.}~\bibnamefont {Musoke}},\
  and\ \bibinfo {author} {\bibfnamefont {C.}~\bibnamefont
  {Prescod-Weinstein}},\ }\href {https://doi.org/10.1103/PhysRevD.107.063520}
  {\bibfield  {journal} {\bibinfo  {journal} {Phys. Rev. D}\ }\textbf {\bibinfo
  {volume} {107}},\ \bibinfo {pages} {063520} (\bibinfo {year} {2023})},\
  \Eprint {https://arxiv.org/abs/2302.04302} {arXiv:2302.04302 [astro-ph.CO]}
  \BibitemShut {NoStop}%
\bibitem [{\citenamefont {Amin}\ \emph {et~al.}(2022)\citenamefont {Amin},
  \citenamefont {Jain}, \citenamefont {Karur},\ and\ \citenamefont
  {Mocz}}]{Amin:2022pzv}%
  \BibitemOpen
  \bibfield  {author} {\bibinfo {author} {\bibfnamefont {M.~A.}\ \bibnamefont
  {Amin}}, \bibinfo {author} {\bibfnamefont {M.}~\bibnamefont {Jain}}, \bibinfo
  {author} {\bibfnamefont {R.}~\bibnamefont {Karur}},\ and\ \bibinfo {author}
  {\bibfnamefont {P.}~\bibnamefont {Mocz}},\ }\href
  {https://doi.org/10.1088/1475-7516/2022/08/014} {\bibfield  {journal}
  {\bibinfo  {journal} {JCAP}\ }\textbf {\bibinfo {volume} {08}}\bibfield
  {number} {\bibinfo  {number} { (08)},\ \bibinfo {pages} {014}},\ }\Eprint
  {https://arxiv.org/abs/2203.11935} {arXiv:2203.11935 [astro-ph.CO]}
  \BibitemShut {NoStop}%
\bibitem [{\citenamefont {{Hogan}}\ and\ \citenamefont
  {{Rees}}(1988)}]{Hogan:1988mp}%
  \BibitemOpen
  \bibfield  {author} {\bibinfo {author} {\bibfnamefont {C.~J.}\ \bibnamefont
  {{Hogan}}}\ and\ \bibinfo {author} {\bibfnamefont {M.~J.}\ \bibnamefont
  {{Rees}}},\ }\href {https://doi.org/10.1016/0370-2693(88)91655-3} {\bibfield
  {journal} {\bibinfo  {journal} {Phys. Lett. B}\ }\textbf {\bibinfo {volume}
  {205}},\ \bibinfo {pages} {228} (\bibinfo {year} {1988})}\BibitemShut
  {NoStop}%
\bibitem [{\citenamefont {{Kolb}}\ and\ \citenamefont
  {{Tkachev}}(1994)}]{Kolb:1994fi}%
  \BibitemOpen
  \bibfield  {author} {\bibinfo {author} {\bibfnamefont {E.~W.}\ \bibnamefont
  {{Kolb}}}\ and\ \bibinfo {author} {\bibfnamefont {I.~I.}\ \bibnamefont
  {{Tkachev}}},\ }\href {https://doi.org/10.1103/PhysRevD.50.769} {\bibfield
  {journal} {\bibinfo  {journal} {\prd}\ }\textbf {\bibinfo {volume} {50}},\
  \bibinfo {pages} {769} (\bibinfo {year} {1994})},\ \Eprint
  {https://arxiv.org/abs/astro-ph/9403011} {astro-ph/9403011} \BibitemShut
  {NoStop}%
\bibitem [{\citenamefont {Ellis}\ \emph {et~al.}(2021)\citenamefont {Ellis},
  \citenamefont {Marsh},\ and\ \citenamefont {Behrens}}]{Ellis:2020gtq}%
  \BibitemOpen
  \bibfield  {author} {\bibinfo {author} {\bibfnamefont {D.}~\bibnamefont
  {Ellis}}, \bibinfo {author} {\bibfnamefont {D.~J.~E.}\ \bibnamefont
  {Marsh}},\ and\ \bibinfo {author} {\bibfnamefont {C.}~\bibnamefont
  {Behrens}},\ }\href {https://doi.org/10.1103/PhysRevD.103.083525} {\bibfield
  {journal} {\bibinfo  {journal} {Phys. Rev. D}\ }\textbf {\bibinfo {volume}
  {103}},\ \bibinfo {pages} {083525} (\bibinfo {year} {2021})},\ \Eprint
  {https://arxiv.org/abs/2006.08637} {arXiv:2006.08637 [astro-ph.CO]}
  \BibitemShut {NoStop}%
\bibitem [{\citenamefont {O'Hare}\ \emph {et~al.}(2022)\citenamefont {O'Hare},
  \citenamefont {Pierobon}, \citenamefont {Redondo},\ and\ \citenamefont
  {Wong}}]{OHare:2021zrq}%
  \BibitemOpen
  \bibfield  {author} {\bibinfo {author} {\bibfnamefont {C.~A.~J.}\
  \bibnamefont {O'Hare}}, \bibinfo {author} {\bibfnamefont {G.}~\bibnamefont
  {Pierobon}}, \bibinfo {author} {\bibfnamefont {J.}~\bibnamefont {Redondo}},\
  and\ \bibinfo {author} {\bibfnamefont {Y.~Y.~Y.}\ \bibnamefont {Wong}},\
  }\href {https://doi.org/10.1103/PhysRevD.105.055025} {\bibfield  {journal}
  {\bibinfo  {journal} {Phys. Rev. D}\ }\textbf {\bibinfo {volume} {105}},\
  \bibinfo {pages} {055025} (\bibinfo {year} {2022})},\ \Eprint
  {https://arxiv.org/abs/2112.05117} {arXiv:2112.05117 [hep-ph]} \BibitemShut
  {NoStop}%
\bibitem [{\citenamefont {Zeng}\ \emph {et~al.}(2025)\citenamefont {Zeng},
  \citenamefont {Zhang},\ and\ \citenamefont {Chen}}]{Zeng:2025unb}%
  \BibitemOpen
  \bibfield  {author} {\bibinfo {author} {\bibfnamefont {Y.}~\bibnamefont
  {Zeng}}, \bibinfo {author} {\bibfnamefont {B.}~\bibnamefont {Zhang}},\ and\
  \bibinfo {author} {\bibfnamefont {J.}~\bibnamefont {Chen}},\ }\href@noop {}
  {\bibinfo {title} {{Self-Interaction Controls Vortex Scale in Soliton
  Mergers}}} (\bibinfo {year} {2025}),\ \Eprint
  {https://arxiv.org/abs/2509.07401} {arXiv:2509.07401 [hep-ph]} \BibitemShut
  {NoStop}%
\bibitem [{\citenamefont {Chen}(2026)}]{Chen:2026sph}%
  \BibitemOpen
  \bibfield  {author} {\bibinfo {author} {\bibfnamefont {J.}~\bibnamefont
  {Chen}},\ }\href@noop {} {\bibinfo {title} {{Yukawa-Screened Bose-Star
  Condensation}}} (\bibinfo {year} {2026}),\ \Eprint
  {https://arxiv.org/abs/2605.23206} {arXiv:2605.23206 [hep-ph]} \BibitemShut
  {NoStop}%
\bibitem [{\citenamefont {Sheth}\ and\ \citenamefont
  {Tormen}(1999)}]{Sheth:1999mn}%
  \BibitemOpen
  \bibfield  {author} {\bibinfo {author} {\bibfnamefont {R.~K.}\ \bibnamefont
  {Sheth}}\ and\ \bibinfo {author} {\bibfnamefont {G.}~\bibnamefont {Tormen}},\
  }\href {https://doi.org/10.1046/j.1365-8711.1999.02692.x} {\bibfield
  {journal} {\bibinfo  {journal} {Mon. Not. Roy. Astron. Soc.}\ }\textbf
  {\bibinfo {volume} {308}},\ \bibinfo {pages} {119} (\bibinfo {year}
  {1999})},\ \Eprint {https://arxiv.org/abs/astro-ph/9901122}
  {arXiv:astro-ph/9901122} \BibitemShut {NoStop}%
\bibitem [{\citenamefont {{Schive}}\ \emph {et~al.}(2016)\citenamefont
  {{Schive}}, \citenamefont {{Chiueh}}, \citenamefont {{Broadhurst}},\ and\
  \citenamefont {{Huang}}}]{2016ApJ...818...89S}%
  \BibitemOpen
  \bibfield  {author} {\bibinfo {author} {\bibfnamefont {H.-Y.}\ \bibnamefont
  {{Schive}}}, \bibinfo {author} {\bibfnamefont {T.}~\bibnamefont {{Chiueh}}},
  \bibinfo {author} {\bibfnamefont {T.}~\bibnamefont {{Broadhurst}}},\ and\
  \bibinfo {author} {\bibfnamefont {K.-W.}\ \bibnamefont {{Huang}}},\ }\href
  {https://doi.org/10.3847/0004-637X/818/1/89} {\bibfield  {journal} {\bibinfo
  {journal} {\apj}\ }\textbf {\bibinfo {volume} {818}},\ \bibinfo {eid} {89}
  (\bibinfo {year} {2016})},\ \Eprint {https://arxiv.org/abs/1508.04621}
  {arXiv:1508.04621} \BibitemShut {NoStop}%
\bibitem [{\citenamefont {Amaral}\ \emph {et~al.}(2024)\citenamefont {Amaral},
  \citenamefont {Jain}, \citenamefont {Amin},\ and\ \citenamefont
  {Tunnell}}]{Amaral_2024}%
  \BibitemOpen
  \bibfield  {author} {\bibinfo {author} {\bibfnamefont {D.~W.}\ \bibnamefont
  {Amaral}}, \bibinfo {author} {\bibfnamefont {M.}~\bibnamefont {Jain}},
  \bibinfo {author} {\bibfnamefont {M.~A.}\ \bibnamefont {Amin}},\ and\
  \bibinfo {author} {\bibfnamefont {C.}~\bibnamefont {Tunnell}},\ }\href
  {https://doi.org/10.1088/1475-7516/2024/06/050} {\bibfield  {journal}
  {\bibinfo  {journal} {Journal of Cosmology and Astroparticle Physics}\
  }\textbf {\bibinfo {volume} {2024}}\bibinfo  {number} { (06)},\ \bibinfo
  {pages} {050}}\BibitemShut {NoStop}%
\end{thebibliography}%

\appendix
\section{Halo and Proca-star definitions}
\label{app:catalog}

{
The total dimensionless vector-field density and overdensity are
\begin{align}
\widetilde\rho(\widetilde{\bm x})
&=\sum_{j=x,y,z}|\widetilde\psi_j(\widetilde{\bm x})|^2,
\label{eq:dimensionless_density}\\
\delta(\widetilde{\bm x})
&=\frac{\widetilde\rho(\widetilde{\bm x})-
\overline{\widetilde\rho}}{\overline{\widetilde\rho}}.
\label{eq:overdensity_def}
\end{align}

We use the overdensity field to construct a halo and Proca star catalogue for analysis as follows. At each output of the halo finding algorithm we locate a density peak, and after counting it in the catalogue, the peak is removed from the density field. At each output we select the largest remaining peak and require
\(\delta_{\rm max}>200\).  Around the accepted peak we construct the
spherically averaged overdensity using periodic minimum-image distances and
define the halo boundary by
\begin{equation}
\delta_{\rm sph}(\widetilde R_{\rm h})=25.
\label{eq:halo_radius_def}
\end{equation}
The enclosed halo mass is
\begin{equation}
\widetilde M_{\rm h}
=\int_{|\widetilde{\bm x}-\widetilde{\bm x}_{\rm p}|<\widetilde R_{\rm h}}
\mathrm d^3\widetilde x\,\widetilde\rho
=\sum_{r_{ijk}<\widetilde R_{\rm h}}
\widetilde\rho_{ijk}(\Delta\widetilde x)^3.
\label{eq:halo_mass_def}
\end{equation}
The assigned region is excluded from subsequent peak searches, and the
procedure is repeated until no peak with \(\delta_{\rm max}>200\) remains.

For each halo center we fit the contiguous inner radial density profile to
\begin{equation}
\widetilde\rho_{\rm fit}(\widetilde r)
=\widetilde\rho_c
\left[1+\left(0.230\frac{\widetilde r}{\widetilde r_s}\right)^2\right]^{-8},
\label{eq:soliton_profile_fit}
\end{equation}
where \(\widetilde\rho_c\) and \(\widetilde r_s\) are fitted independently.
This density corresponds to the wavefunction-amplitude shape
\(\widetilde\psi_{\rm sol}\propto\widetilde r_s^{-2}
[1+(0.230\widetilde r/\widetilde r_s)^2]^{-4}\)
\cite{Gorghetto:2022sue}.  The surrounding fuzzy-halo envelope is excluded.
The profile agreement is quantified by
\begin{equation}
\epsilon_{\rm prof}=\left\{\frac{1}{N_{\rm fit}}\sum_i
\left[\log_{10}\widetilde\rho_i-
\log_{10}\widetilde\rho_{\rm fit}(\widetilde r_i)\right]^2\right\}^{1/2}.
\label{eq:profile_fit_rmse}
\end{equation}
We retain profiles with \(\epsilon_{\rm prof}<0.5\) dex, at least six
independent radial samples, and a fitted density dynamic range of at least
0.5 dex.  Halos that fail these profile conditions remain in the halo sample 
but do not enter the Proca-star core--halo relation.

For each accepted profile, the Proca-star mass \(\widetilde M_\star\) is
defined by integrating only over the radial interval participating in the
fit,
\begin{equation}
\widetilde M_\star
=4\pi\int_0^{\widetilde R_{\rm fit}}\mathrm d\widetilde r\,
\widetilde r^2\widetilde\rho_{\rm fit}(\widetilde r),
\label{eq:soliton_mass_profile}
\end{equation}
where \(\widetilde R_{\rm fit}\) is the outermost radius of the accepted
contiguous interval.  No fitted-profile extrapolation into the surrounding
halo is included.  Both masses are therefore measured in the same simulation
mass units.  All systems satisfying the profile conditions are included.
}

\end{document}